\documentclass[reprint,onecolumn,superscriptaddress,amsmath,amssymb,aps,pre,longbibliography,nofootinbib]{revtex4-2}

\usepackage{enumitem}
\usepackage{amsmath}
\usepackage{mathtools}
\usepackage{graphicx}
\usepackage{bm}
\usepackage[dvipsnames]{xcolor}
\usepackage[colorlinks=true,allcolors=red!50!black]{hyperref}
\usepackage{microtype}

\DeclareMathOperator{\var}{var}
\DeclareMathOperator{\cov}{cov}
\DeclareMathOperator{\sech}{sech}
\DeclareMathOperator{\csch}{csch}

\begin{document}

\title{A note on the dynamics of extended-context disordered kinetic spin models}

\author{Jacob A. Zavatone-Veth}
\email{jzavatoneveth@fas.harvard.edu}
\affiliation{Society of Fellows, Harvard University, Cambridge, MA, USA}
\affiliation{Center for Brain Science, Harvard University, Cambridge, MA, USA}

\author{Cengiz Pehlevan}
\email{cpehlevan@seas.harvard.edu}
\affiliation{Center for Brain Science, Harvard University, Cambridge, MA, USA}
\affiliation{John A. Paulson School of Engineering and Applied Sciences, Harvard University, Cambridge, MA, USA}
\affiliation{Kempner Institute for the Study of Natural and Artificial Intelligence, Harvard University, Cambridge, MA, USA}

\begin{abstract}%
    Inspired by striking advances in language modeling, there has recently been much interest in developing autogressive sequence models that are amenable to analytical study. In this short note, we consider extensions of simple disordered kinetic glass models from statistical physics. These models have tunable correlations, are easy to sample, and can be solved exactly when the state space dimension is large. In particular, we give an expository derivation of the dynamical mean field theories that describe their asymptotic statistics. We therefore propose that they constitute an interesting set of toy models for autoregressive sequence generation, in which one might study learning dynamics. 
\end{abstract}

\date{\today}

\maketitle

\section{Introduction}

There has recently been substantial interest in autoregressive sequence modeling, \textit{i.e.}, in models for probability distributions of the form $p(\mathbf{s}_{t} \,|\, \mathbf{s}_{t-1},\ldots,\mathbf{s}_{t-K})$ for some tokens $\mathbf{s}_{t}$ and a context length $K$, as modern large language models (LLMs) are fundamentally of this form \cite{radford2018improving,brown2020gpt3,openai2023gpt4}. Given the striking capabilities of such models, developing a theoretical understanding of how those abilities depend on data structure and model architecture is a pressing goal. From the perspective of the statistical physics of learning, we would like to devise a setting in which we could study how a `student' model learns to imitate sequences generated by a `teacher' of similar architecture \cite{engel2001statistical,lu2025asymptotic}. The key challenge is to formulate an interesting yet analytically tractable class of models for distributions of this form. Recent work by \citet{rende2024optimal} has shown that a modified Potts model is a suitable candidate for a teacher in the context of \emph{masked} sequence modeling, but the autoregressive setting is still open. 

What are the desiderata that a candidate toy autoregressive sequence model should satisfy? We propose that such a model should be analytically tractable, at least in some limits, be easy to sample numerically, display interesting statistical properties, e.g. tunable temporal correlations, and resemble `real-world' sequence models, either in its structure or in its statistical properties. In this note, we take inspiration from statistical physics \cite{crisanti1987spherical,crisanti1988ising,aguilera2021unifying,aguilera2023nonequilibrium}, to propose a simple model satisfying these four desiderata. 

\section{Inspiration: the kinetic Sherrington-Kirkpatrick model}

From a physical perspective, a natural starting point is the fully asymmetric kinetic Sherrington-Kirkpatrick (kSK) model. The kSK model is a Markov chain on $\{-1,+1\}^{N}$ with 
\begin{align}
    p_{\mathbf{J}}(\mathbf{s}_{t}\,|\,\mathbf{s}_{t-1}) = \frac{\exp[-\beta \mathbf{s}_{t}^{\top} \mathbf{J} \mathbf{s}_{t-1} ]}{\sum_{\mathbf{u}_{t} \in \{-1,+1\}^{N}} \exp[-\beta \mathbf{u}_{t}^{\top} \mathbf{J} \mathbf{s}_{t-1}]}
\end{align}
for an inverse temperature $\beta > 0$ and a quenched random interaction matrix $\mathbf{J}$ with i.i.d. Gaussian elements $J_{ij} \sim \mathcal{N}(0, g^{2}/N)$. For simplicity, we will focus on the `pure' SK case, where the interactions have mean zero and there is no external magnetic field to bias the state distribution.

After being studied by \citet{crisanti1987spherical,crisanti1988ising} in the late 1980s for various continuous-time dynamics, the discrete-time version of the kSK model has recently emerged as a paradigmatic model for the thermodynamics of nonequilibrium complex systems \cite{aguilera2021unifying,aguilera2023nonequilibrium}. Moreover, \citet{bal2023spin} has observed in a series of blog posts that the kSK model bears a strong structural resemblance to the Attention mechanism used in modern autoregressive language models. At large $N$ the kSK model is amenable to solution by mean-field techniques, and it is easy to sample thanks to the fact that the sites are conditionally independent given the local field $\mathbf{h}_{t} = - \beta \mathbf{J} \mathbf{s}_{t-1}$, \textit{i.e.}, we have the factorization
\begin{align}
    p_{\mathbf{J}}(\mathbf{s}_{t}\,|\,\mathbf{s}_{t-1}) = \prod_{j = 1}^{N} \frac{\exp[s_{t,j} h_{t,j}]}{2 \cosh(h_{t,j})}.
\end{align}
However, this model is insufficient for our purposes, because it exhibits single-step decorrelation, \textit{i.e.}, $\mathbb{E}_{\mathbf{J}} \langle \mathbf{s}_{t} \cdot \mathbf{s}_{t+1} \rangle = 0$ for all times $t$, where $\langle \cdot \rangle$ denotes averaging over the random process for some fixed initial condition $\mathbf{s}_{0}$ and a fixed realization of $\mathbf{J}$ for a quantity depending on times up to $T$. This decorrelation follows from a simple symmetry argument. Concretely, we observe that the transformation $\tilde{\mathbf{s}}_{t} = - \mathbf{s}_{t}$ for $t$ odd, $\tilde{\mathbf{s}}_{t} = \mathbf{s}_{t}$ for $t$ even, and $\tilde{\mathbf{J}} = - \mathbf{J}$ leaves the combined measure in $\mathbb{E}_{\mathbf{J}} \langle \cdot \rangle$ invariant for any fixed $\mathbf{s}_{0}$. This implies immediately that $\mathbb{E}_{\mathbf{J}} \langle \mathbf{s}_{t} \rangle = \mathbf{0}$ for any odd $t$, and that $\mathbb{E}_{\mathbf{J}} \langle \mathbf{s}_{t} \cdot \mathbf{s}_{t'} \rangle = 0$ whenever $t$ and $t'$ have opposite parity. The reason for the structure of this transformation is that the $\mathbf{s}_{0}$-dependent term in the measure is invariant under flipping the sign of $\mathbf{J}$ only if $\mathbf{s}_{1}$ is also negated. This argument implies that one always has one-step decorrelation for this model, \textit{i.e.}, $\mathbb{E}_{\mathbf{J}} \langle \mathbf{s}_{t} \cdot \mathbf{s}_{t+1} \rangle = 0$. Single-step decorrelation could be avoided by adding a positive mean to $\mathbf{J}$, so as to encourage alignment between $\mathbf{s}_{t}$ and $\mathbf{s}_{t-1}$, but this does not yield flexibly tunable temporal correlations \cite{aguilera2023nonequilibrium}. We will therefore pursue an alternative approach. 

\section{Construction of the model}

We now introduce the three classes of extended-context models we consider in this work. Consider an $N$-dimensional state space $\mathcal{S}_{N}$, which we conceptualize as representing the token embedding space of a sequence model. We will take $N$ to be large, consistent with the use of embedding spaces of dimension greater than 10,000 in modern sequence models \cite{brown2020gpt3}. Equip $\mathcal{S}_{N}$ with a (possibly un-normalized) probability measure $\sigma_{N}$ that is reflection-symmetric, \textit{i.e.}, invariant under $\mathbf{s} \mapsto - \mathbf{s}$ for all $\mathbf{s} \in \mathcal{S}_{N}$. We consider three models:
\begin{enumerate}[leftmargin=*,noitemsep]
    \item \textbf{Ising}: 
    $\mathcal{S}_{N} = \{-1,+1\}^{N}$ with $\sigma_{N}$ the uniform counting measure. 

    \item \textbf{Gaussian}: with $\mathcal{S}_{N} = \mathbb{R}^{N}$ and $\sigma_{N}$ the standard Gaussian measure

    \item \textbf{Spherical}: $\mathcal{S}_{N} = \{\mathbf{s} \in \mathbb{R}^{N} : \Vert \mathbf{s} \Vert_{2}^{2} = N\}$ and $\sigma_{N}$ the uniform probability measure on the sphere. 
    
\end{enumerate}
Define a Markov chain of order $K$ on $\mathcal{S}_{N}$ by the transition probabilities with density 
\begin{equation}
    p_{\{\mathbf{J}_{k}\}}(\mathbf{s}_{t}\, |\, \mathbf{s}_{t-1},\ldots \mathbf{s}_{t-K}) = \frac{\exp[-\beta E_{\{\mathbf{J}_{k}\}}(\mathbf{s}_{t}; \mathbf{s}_{t-1},\ldots,\mathbf{s}_{t-K})]}{\int d\sigma_{N}(\mathbf{u}_{t})\, \exp[-\beta E_{\{\mathbf{J}_{k}\}}(\mathbf{u}_{t}; \mathbf{s}_{t-1},\ldots,\mathbf{s}_{t-K})]}
\end{equation}
with respect to $\sigma_{N}$, for an inverse temperature $\beta > 0$ and an energy function
\begin{align}
    E_{\{\mathbf{J}_{k}\}}(\mathbf{s}_{t}; \mathbf{s}_{t-1},\ldots,\mathbf{s}_{t-K}) = \sum_{k=1}^{K} \mathbf{s}_{t}^{\top} \mathbf{J}_{k} \mathbf{s}_{t-k}  
\end{align}
for a set of interaction matrices $\{\mathbf{J}_{k}\}_{k=1}^{K}$. Again, we write $\langle \cdot \rangle$ for the average over the random process for some fixed initial sequence $\mathbf{s}_{0}, \ldots, \mathbf{s}_{1-K}$, \textit{i.e.}, for a quantity depending on times up to $T$, we have 
\begin{align}
    \langle \cdot \rangle = \int \prod_{t=1}^{T} d\sigma_{N}(\mathbf{s}_{t})\, (\cdot) \prod_{t=1}^{T} p_{\{\mathbf{J}_{k}\}}(\mathbf{s}_{t}\, |\, \mathbf{s}_{t-1},\ldots \mathbf{s}_{t-K}).
\end{align}

We generalize the fully asymmetric kSK model by taking the interaction matrices $\mathbf{J}_{k}$ to be jointly Gaussian, with zero mean. As a concrete model, we will focus on the case in which the interactions are uncorrelated across sites but possibly correlated across lags, \textit{i.e.}, 
\begin{align}
    \mathbb{E}_{\mathbf{J}}[ (J_{k})_{ij} (J_{k'})_{i'j'}] = \frac{1}{N} \delta_{ii'} \delta_{jj'} \Gamma_{k,k'} 
\end{align}
for some correlation matrix $\Gamma_{k,k'}$.\footnote{As in related random matrix problems \cite{zv2023wishart}, one could in principle further generalize the analysis to allow non-trivial correlations between the elements of each row of the interaction matrix $\mathbf{J}_{k}$---\textit{i.e.}, to replace $\delta_{jj'}$ with some non-diagonal matrix---but here we focus on the minimal setting with spatially uncorrelated interactions.} Our objective is to study the resulting ensemble of random dynamical systems with quenched disorder.

For any context length, all of these models are easy to sample from numerically. In the Ising case, as for the simple kinetic SK model, we have independence of sites conditioned on the local field, as $p_{\{\mathbf{J}_{k}\}}(\mathbf{s}_{t}\, |\, \mathbf{s}_{t-1},\ldots \mathbf{s}_{t-K}) = \prod_{j=1}^{N} \frac{\exp( s_{t,j} h_{t,j} )}{2 \cosh( \beta h_{t,j} )}$ for $\mathbf{h}_{t} = - \beta \sum_{k=1}^{K} \mathbf{J}_{k} \mathbf{s}_{t-k}$. In the Gaussian case, we have $p_{\{\mathbf{J}_{k}\}}(\mathbf{s}_{t}\, |\, \mathbf{s}_{t-1},\ldots \mathbf{s}_{t-K}) \,d\sigma_{N}(\mathbf{s}_{t}) \propto \exp\left[-\frac{1}{2} \Vert \mathbf{s}_{t} - \mathbf{h}_{t} \Vert^{2} \right] \,d\mathbf{s}_{t}$ up to normalization, so here again sites are independent conditioned on the local field. In the spherical case, we see immediately from the expression $p_{\{\mathbf{J}_{k}\}}(\mathbf{s}_{t}\, |\, \mathbf{s}_{t-1},\ldots \mathbf{s}_{t-K}) \propto \exp[ \mathbf{h}_{t}^{\top} \mathbf{s}_{t}]$ that, conditioned on the context, $\mathbf{s}_{t}$ follows a von Mises-Fisher distribution with mean direction $\bm{\mu} = -\frac{\sqrt{N} \mathbf{h}_{t}}{\Vert \mathbf{h}_{t} \Vert}$ and concentration parameter  $\kappa = \frac{\Vert \mathbf{h}_{t} \Vert}{N}$ on the $N$-sphere of radius $\sqrt{N}$. This is easy to sample using standard algorithms \cite{wood1994vmf}. Therefore, all three models satisfy the desideratum of efficient sampling.

\section{Dynamical mean-field theory}

We now turn to the questions of whether these models are analytically tractable, and if they display interesting statistical behavior. In the limit where $N \to \infty$ for fixed $T$ and $K$, the asymptotic statistics of these models are easy to characterize using standard dynamical field theory (DMFT) approaches from the statistical physics of disordered systems. In particular, we consider the generating function 
\begin{align}
    Z[\{\mathbf{b}_{t}\} ; \{\mathbf{J}_{k}\}] = \left\langle \exp\left[\sum_{t=1}^{T} \mathbf{b}_{t} \cdot \mathbf{s}_{t} \right] \right\rangle .
\end{align}
As usual \cite{crisanti1987spherical,crisanti1988ising}, the fact that $Z[\{\mathbf{b}_{t}\} ; \{\mathbf{J}_{k}\}] |_{\{\mathbf{b}_{t} = \mathbf{0}\}} = 1$ for any $\{\mathbf{J}_{k}\}$ implies that the generating function can be averaged directly over the quenched disorder to obtain $Z[\{\mathbf{b}_{t}\} ] = \mathbb{E}_{\mathbf{J}}Z[\{\mathbf{b}_{t}\} ; \{\mathbf{J}_{k}\}]$, from which we can immediately compute quenched moments, \textit{e.g.}, $\mathbb{E}_{\mathbf{J}} \langle \mathbf{s}_{t} \rangle = {\partial} Z[\{\mathbf{b}_{t}\} ] /{\partial \mathbf{b}_{t}}|_{\{\mathbf{b}_{t} = \mathbf{0}\}}$. 

In the Appendices, we give an expository derivation of the DMFT equations for each variant of the models we consider. For all three models, the DMFT order parameter is the self-averaging two-point function
\begin{align}
    C_{t,t'} = \mathbb{E}_{\mathbf{J}} \left\langle \frac{1}{N} \mathbf{s}_{t} \cdot \mathbf{s}_{t'} \right\rangle, 
\end{align}
in terms of which the effective single-site distribution is self-consistently determined. In the Ising-like and spherical models, the Cauchy-Schwarz inequality and triangle inequality together imply that $|C_{t,t'}| \leq 1$, while in the Gaussian case boundedness is not guaranteed. For all three models, for $t,t' = 1-K, \ldots, 0$, $C_{t,t'}$ is entirely fixed by the initial condition as $C_{t,t'} = \frac{1}{N} \sum_{j=1}^{N} s_{t,j} s_{t',j}$, while for $t = 1, \ldots, T$ and $t' = 1-K, \ldots, 0$ we have $C_{t,t'} = 0$. For brevity, let
\begin{align} \label{eqn:sigma}
    \Sigma_{t,t'} = \beta^{2} \sum_{k,k'=1}^{K} \Gamma_{k,k'} C_{t-k,t'-k'}. 
\end{align}
We can now state the results of the DMFT analysis.

\begin{figure}[t]
    \centering
    \includegraphics[width=7in]{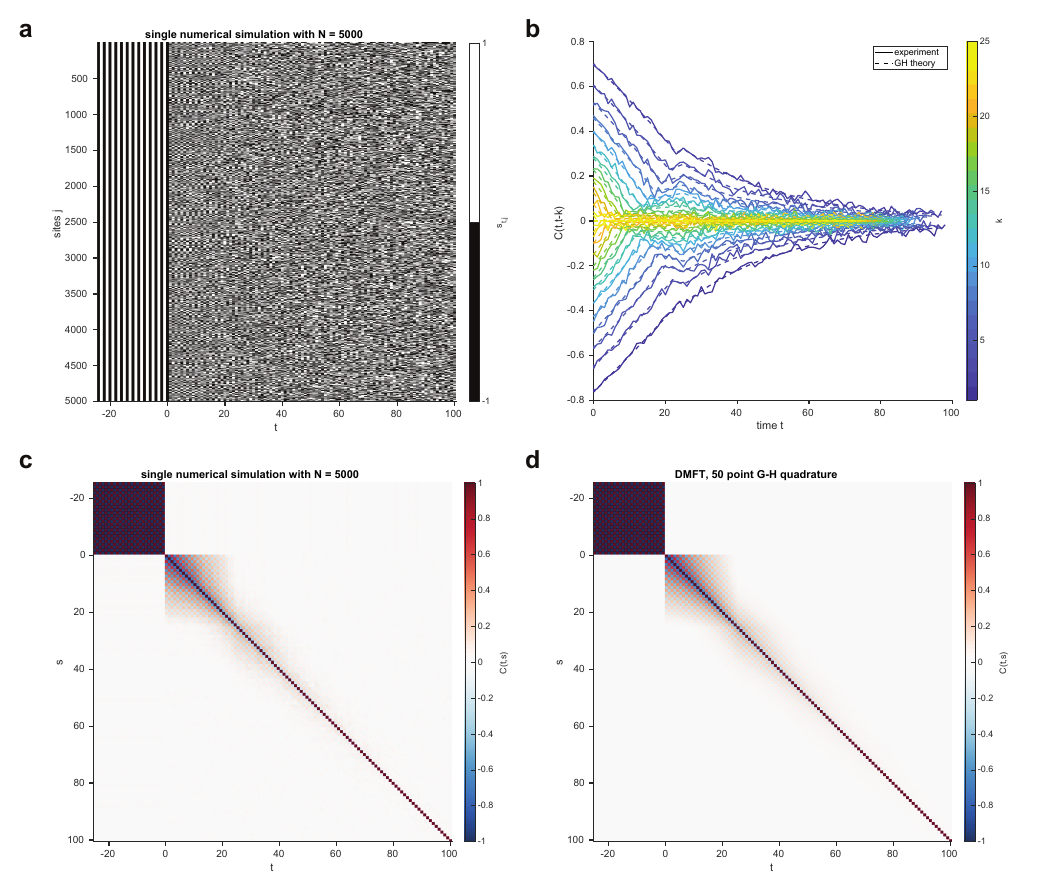}
    \caption{Simulation of an Ising-like model with $\Gamma_{k,k'} = \delta_{k,k'}$ and $K=25$, with $N=5000$. \textbf{a}. State $\mathbf{s}_{t}$ over time. Times before $t=0$ represent the initial condition, which is chosen arbitrarily. \textbf{b}. Slices through the DMFT autocorrelation function $C_{t,t-k}$ across time for varying lags $k$, showing that the DMFT accurately predicts the empirically-measured correlation from a single simulation. The expectation in the DMFT equations is numerically evaluated using 50-point Gauss-Hermite quadrature. We see that autocorrelations at all non-zero lags decay over time. \textbf{c}. The DMFT autocorrelation function $C_{t,t'}$ from a single numerical simulation for which the slices are shown at top right. \textbf{d}. The corresponding DMFT prediction for the autocorrelation function. }
    \label{fig:ising_uncorrelated}
\end{figure}

\subsection{Do we need correlated weights?}

\begin{figure}
    \centering
    \includegraphics[width=7in]{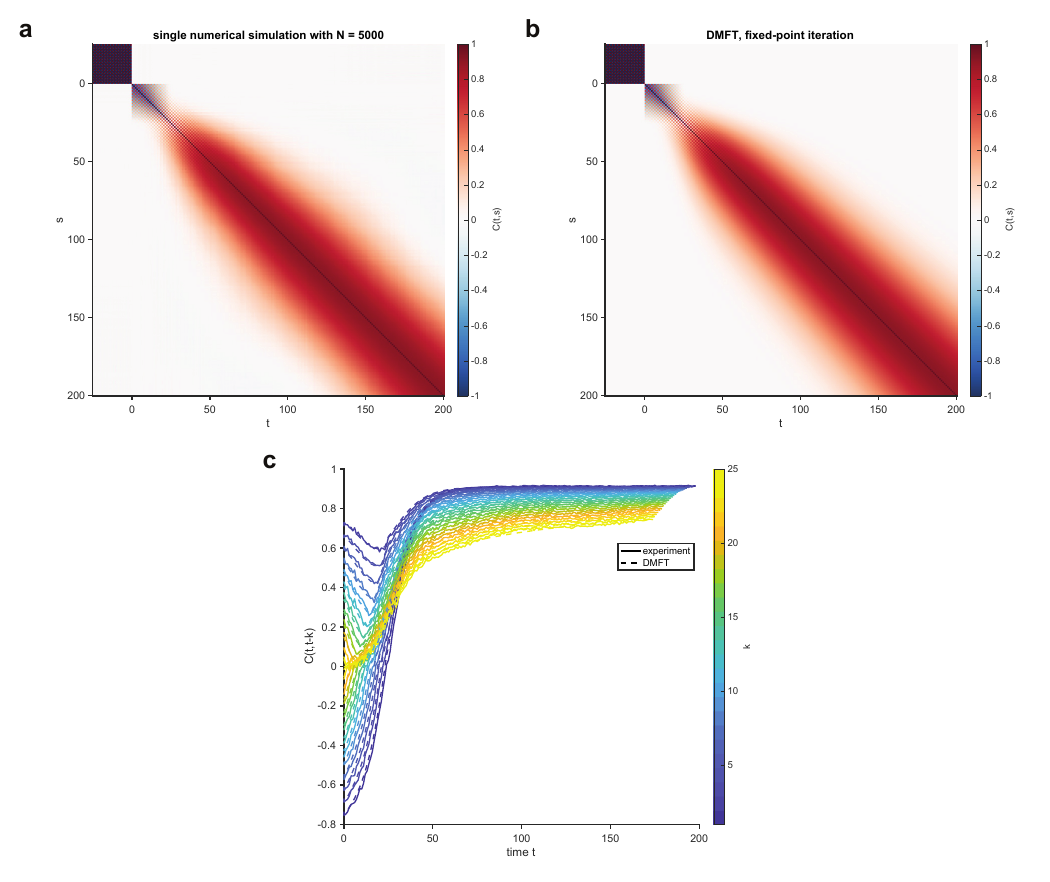}
    \caption{Simulations of a spherical model with $\Gamma_{k,k'} = \delta_{k,k'} + r (1-\delta_{k,k'})$ for $r = 0.25$  and $K=25$. Here, $\beta = 1$. The top row shows heatmaps of the empirical (\textbf{a}) and DMFT (\textbf{b}) correlation functions $C_{t,s}$. The bottom row shows slices through the correlation functions (\textbf{c}). }
    \label{fig:spherical_model_equicorrelated}
\end{figure}

\begin{figure}
    \centering
    \includegraphics[width=7in]{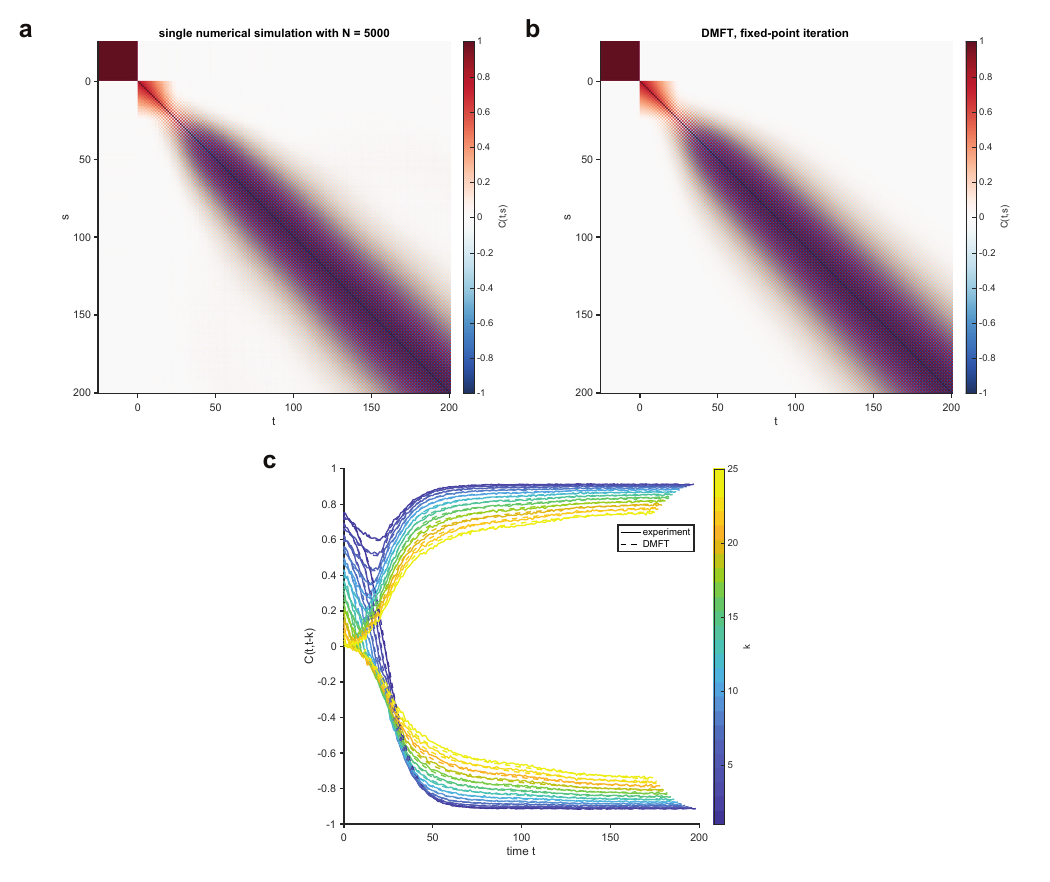}
    \caption{Simulations of a spherical model with $\Gamma_{k,k'} = \delta_{k,k'} + r (1-\delta_{k,k'}) (-1)^{k+k'}$ for $r = 0.25$  and $K=25$. Here, $\beta = 1$. The top row shows heatmaps of the empirical (\textbf{a}) and DMFT (\textbf{b}) correlation functions $C_{t,s}$. The bottom row shows slices through the correlation functions (\textbf{c}). }
    \label{fig:spherical_model_osc}
\end{figure}

Before presenting the details of the DMFT descriptions of each variant of the model at hand, we briefly address the question of what we gain from introducing correlations to the interaction distribution, as the answer for all three model types is the same. In all variants of the model, the DMFT is a causal iterative equation for $C_{t,t'}$, with all dependence on previous times being through $\Sigma_{t,t'}$ as defined in \eqref{eqn:sigma}. We are interested mostly in the long-time properties of these DMFTs, particularly when they admit stationary solutions of the form $C_{t,t'} = C_{t-t'}$ at long times $t,t' \gg 1$. A natural question at this point is whether we really need correlations in weights across lags to get a non-trivial stationary state, \textit{i.e.}, whether we could take $\Gamma_{k,k'} = \gamma_{k} \delta_{k,k'}$ and get interesting non-vanishing temporal correlation at long times. As anticipated by the fact that we included correlations in the setup, the answer to this question is no. With $\Gamma_{k,k'} = \gamma_{k} \delta_{k,k'}$, we have $\Sigma_{t,t'} = \beta^{2} \sum_{k=1}^{K} \gamma_{k} C_{t-k,t'-k}$. Under the assumption of stationarity, $C_{t-k,t'-k} = C_{(t-k) - (t'-k)} = C_{t-t'}$, hence $\Sigma_{t,t'} = ( \beta^{2} \sum_{k=1}^{K} \gamma_{k} ) C_{t-t'}$. This shows that correlations across lags are required to obtain non-trivial stationary correlations. We demonstrate this for a few arbitrarily-chosen example choices of $\Gamma_{k,k'}$ in Figures \ref{fig:ising_uncorrelated}, \ref{fig:ising_correlated}, \ref{fig:gaussian_model}, \ref{fig:spherical_model_equicorrelated}, and \ref{fig:spherical_model_osc}, which also show that the DMFT predictions are accurate.

\subsection{Ising}

\begin{figure}[t]
    \centering
    \includegraphics[width=7in]{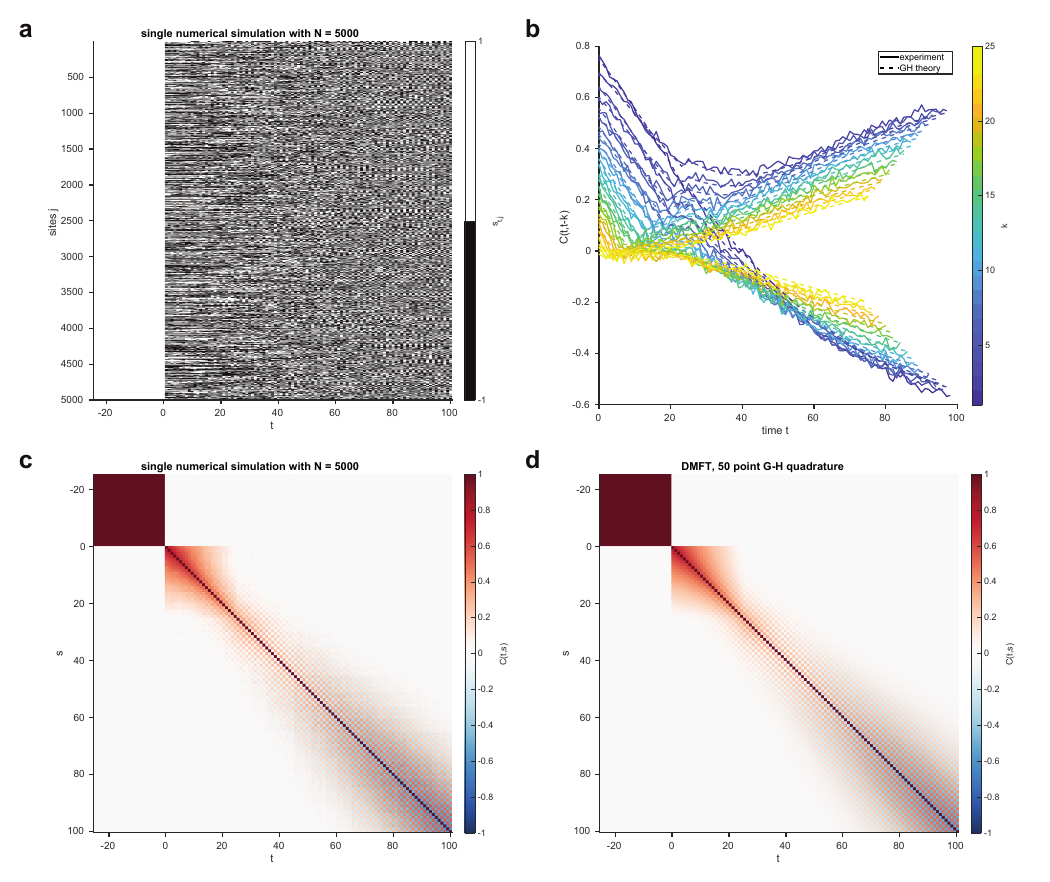}
    \caption{As in Figure \ref{fig:ising_uncorrelated}, but for an Ising-type model with correlated weights $\Gamma_{k,k'} = \delta_{k,k'} + r (1-\delta_{k,k'}) (-1)^{k+k'}$ for $r = 0.1$. Unlike for uncorrelated wights in Figure \ref{fig:ising_uncorrelated}, the autocorrelation does not decay to zero over time. }
    \label{fig:ising_correlated}
\end{figure}

In the Ising case, we show in Appendix \ref{app:ising} that the DMFT equation is
\begin{align}
    C_{t,t'} = \mathbb{E}_{\mathbf{h} \sim \mathcal{N}(\mathbf{0},\bm{\Sigma})}[\tanh(h_{t})\tanh(h_{t'})]
\end{align}
for distinct times $t,t'=1,\ldots,T$ and $C_{t,t} = 1$. Given the fact that $\Sigma_{t,t'}$ depends on the covariance only at previous timesteps \eqref{eqn:sigma}, this recurrence can easily be numerically solved forward in time. The only bottleneck is numerical evaluation of the expectation, which can be performed easily using Gauss-Hermite quadrature, at least at high temperatures (see Figures \ref{fig:ising_uncorrelated} and \ref{fig:ising_correlated}). Here, the single-site generating function is
\begin{align}
    z_{j} = \sum_{\{s_{t,j}\}_{t=1}^{T}} \mathbb{E}_{\mathbf{h}_{j}} \exp\left[\sum_{t=1}^{T} b_{t,j} s_{t,j} \right] \prod_{t=1}^{T} \frac{\exp(s_{t,j} h_{t,j} )}{ 2 \cosh( h_{t,j}) }
\end{align}
for $\mathbf{h}_{j} \sim \mathcal{N}(\mathbf{0},\bm{\Sigma})$, which describes an Ising system in a self-consistently determined field.

\subsection{Gaussian}

\begin{figure}
    \centering
    \includegraphics[width=7in]{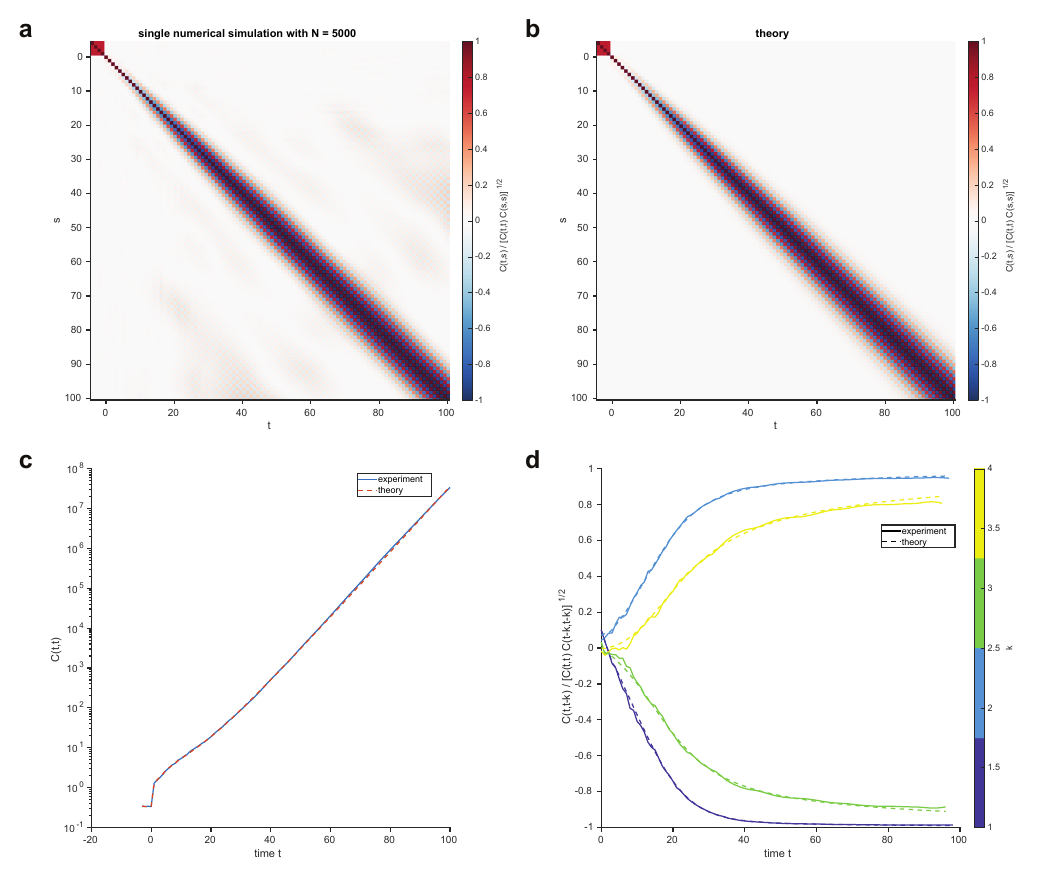}
    \caption{Simulations of a Gaussian model with $\Gamma_{k,k'} = \delta_{k,k'} + r (1-\delta_{k,k'}) (-1)^{k+k'}$ for $r = 0.1$  and $K=4$. Here, $\beta = 0.5$. The top row shows heatmaps of the normalized empirical (\textbf{a}) and DMFT (\textbf{b}) correlation functions $C_{t,s}/\sqrt{C(t,t)C(s,s)}$. The bottom row shows the exponential growth of $C(t,t)$ (\textbf{c}) and slices through the normalized correlation functions (\textbf{d}). }
    \label{fig:gaussian_model}
\end{figure}

In the Gaussian case, we show in Appendix \ref{app:gaussian} that we have the DMFT equation 
\begin{align}
    C_{t,t'} = \delta_{t,t'} + \Sigma_{t,t'}
\end{align}
for $t,t' = 1, \ldots, T$. For the Gaussian model, the single-site generating functions are simply
\begin{align}
    z_{j} = \exp\left[ \frac{1}{2} \sum_{t,t'=1}^{T} C_{t,t'} b_{t,j} b_{t',j} \right].
\end{align}
Again, the causal structure means that the DMFT can be solved forward in time. However, this is now a linear system, which allows greater analytical tractability. That said, the Gaussian model is potentially unstable: if the inverse temperature $\beta$ is too large then the equal-time correlations $C_{t,t}$ grow exponentially with time (Figure \ref{fig:gaussian_model}). We will return to this issue, and to sufficient conditions for stability, in Section \ref{sec:stationary}. 

\subsection{Spherical}

The development of the DMFT for the spherical model is somewhat more complicated than that for the Ising and Gaussian models due to the global constraint on the norm of the state vector. We show in Appendix \ref{app:spherical} how one can derive DMFT equations for this model by applying the replica trick to the normalization terms in the sequence distribution at each timestep. This leads to a DMFT in terms of $C_{t,t'}$ and a set of positive scalars $Q_{t}$, where $C_{t,t'}$ satisfies the recursive equation
\begin{align}
    C_{t,t'} = \frac{1}{Q_{t} Q_{t'}} ( Q_{t} \delta_{t,t'} + \Sigma_{t,t'} ) 
\end{align}
for $t,t' = 1, \ldots, T$, and $Q_{t}$ is determined by the self-consistency condition $C_{t,t} = 1$. One finds that the single-site generating functions are in this case $z_{j} = \exp[ \frac{1}{2} \sum_{t,t'=1}^{T} C_{t,t'} b_{t,j} b_{t',j} ]$, meaning that the fluctuations are Gaussian. These equations are naturally solved via fixed point iteration. For a fixed $Q_{t}$, this is a linear recurrence for $C_{t,t'}$ which can be solved forward in time. Then, we can compute $Q_{t}$ from the given $C_{t,t'}$ using the constraint $C_{t,t} = 1$. Physically, the scalars $Q_{t}$ are up to a shift the Lagrange multipliers that enforce the norm constraint.

\section{Stationary solutions for translation-invariant correlations}\label{sec:stationary}

As uncorrelated weights do not suffice, we now consider whether we can make analytical progress in solving the DMFT equations in the stationary state for translation-invariant correlations of the form
\begin{align}
    \Gamma_{k,k'} = \Gamma_{k-k'} .
\end{align}
By symmetry, we of course have $\Gamma_{k} = \Gamma_{-k}$, where $1-K \leq k \leq K-1$. With this choice, writing $\tau = t - t'$ and $k = k-k'$, we have
\begin{align}
    \Sigma_{\tau} = \beta^{2} \sum_{k = 1-K}^{K-1} (K-|k|) \Gamma_{k} C_{\tau-k}
\end{align}
In the Gaussian and spherical cases, this means that the DMFT equations become a linear convolution equation for $C_{t-t'} = c_{\tau}$: 
\begin{align} \label{eqn:stationary_toeplitz}
    c_{\tau} = \frac{1}{q} \delta_{\tau,0} + \frac{\beta^2}{q^2} \sum_{k= 1-K}^{K-1} (K-|k|) \Gamma_{k} c_{\tau-k}. 
\end{align}
In the Gaussian case, the variable $q$ in \eqref{eqn:stationary_toeplitz} is fixed to $q=1$. In the spherical case, we should have $Q_{t} = q$ assuming stationarity, and we must determine $q$ by imposing the self-consistency condition $c_{\tau=0}=1$ on the solution to \eqref{eqn:stationary_toeplitz}.

The fact that this equation is linear makes the Gaussian and spherical models more amenable to analytical study than their Ising-type counterpart. In particular, this is a banded symmetric Toeplitz system for $c_{\tau}$, which we can solve using the $z$-transform. Before studying the solutions to \eqref{eqn:stationary_toeplitz}, we remark that we may define the inverse temperature $\beta$ such that $\Gamma_{0} = 1$. Then, noting that if $\beta = 0$ we have $c_{\tau=0} = 1/q$, in this limit we should have $q = 1$. This will allow us to sanity-check our general solution.

\subsection{Warm-up: solution for \texorpdfstring{$K=1$}{K=1}}

As a warm-up, if $K = 1$, then \eqref{eqn:stationary_toeplitz} reduces to 
\begin{align}
    c_{\tau} = \frac{1}{q} \delta_{\tau,0} + \frac{\beta^{2}}{q^2}c_{\tau}
\end{align}
which has solution 
\begin{align} \label{eqn:k1_solution}
    c_{\tau} = \begin{cases}
        q (q^{2}-\beta^{2})^{-1} & \tau = 0 \\ 
        0 & \mathrm{otherwise}. 
    \end{cases}
\end{align}
In the Gaussian case, this equation is stable if $\beta < 1$. In the spherical case, imposing the condition that $c_{\tau=0}=1$, we find that $q = \frac{1+\sqrt{1+4\beta^2}}{2}$. Together, this recovers our previous result that with $K=1$ there must be single-step decorrelation. 

\subsection{Stationary solutions for \texorpdfstring{$K>1$}{K>1}}

For $K>1$, taking a $z$-transform
\begin{align}
    C(z) = \sum_{\tau=-\infty}^{\infty} z^{-\tau} c_{\tau}
\end{align}
for $|z| \leq 1$ leads to the equation
\begin{align}
    C(z) = \frac{1}{q} + \frac{\beta^{2}}{q^2} \sum_{k=1-K}^{K-1} (K-|k|) \Gamma_{k} z^{-k} C(\theta) ,
\end{align}
which has formal solution
\begin{align}
    C(z) 
    &= \frac{q}{q^2-\beta^{2} \sum_{k=1-K}^{K-1} (K-|k|) \Gamma_{k} z^{-k}}
    \\
    &= \frac{q}{q^2-K \beta^{2} - \beta^{2} \sum_{k=1}^{K-1} (K-k) \Gamma_{k} (z^{k}+z^{-k}) },
\end{align}
using the fact that $\Gamma_{-k} = \Gamma_{k}$. Then, we can formally obtain the solution $c_{\tau}$ by taking the inverse transform: 
\begin{align} \label{eqn:toeplitz_solution}
    c_{\tau} = \int_{-\pi}^{\pi} \frac{d\theta}{2\pi}  e^{i \tau \theta} C(e^{i\theta}) = \int_{-\pi}^{\pi} \frac{d\theta}{2\pi}  e^{i \tau \theta}\frac{q}{q^2-K \beta^{2} - 2 \beta^{2} \sum_{k=1}^{K-1} (K-k) \Gamma_{k} \cos(k\theta) }.
\end{align}

We can see that, if the denominator of \eqref{eqn:toeplitz_solution} does not vanish, then the system will be stable in the sense that $|c_{\tau}|<\infty$ for all $\tau$. In the spherical case, this stability condition is satisfied thanks to the fact that $q$ must be self-consistently determined so as to make the system stable. In the Gaussian case, where $q=1$, we must check that $\beta$ and $\Gamma_{k}$ are chosen so as to make the system stable. By the triangle inequality, a sufficient condition for stability is that 
\begin{align}
    K \beta^{2} + 2 \beta^{2} \sum_{k=1}^{K-1} (K-k) |\Gamma_{k} | < 1, 
\end{align}
where we recall that we have defined $\beta^2$ such that $\Gamma_{0}=1$. One interpretation of this condition is that it gives a conservative estimate of the maximum inverse temperature allowed in the Gaussian model: 
\begin{align}
    \beta < \frac{1}{\sqrt{K + 2 \sum_{k=1}^{K-1} (K-k) |\Gamma_{k}|}}.
\end{align}
If the temperature is too low, the norm of the state vector may diverge, as shown in the example of Figure \ref{fig:gaussian_model}. There, $K=25$ and $\Gamma_{k} = \delta_{k,0} + \frac{1}{4} (1-\delta_{k,0})$, so $K+2 \sum_{k=1}^{K-1} (K-k) |\Gamma_{k}| = 175$. Thus, the crude estimate above indicates that $\beta < 1/\sqrt{175} \simeq 0.08$ is sufficient to ensure stability; in Figure \ref{fig:gaussian_model} we used $\beta = 1$, which evidently exceeds the stability threshold.

For general context lengths $K>1$ and correlations $\Gamma_{k}$, the integral \eqref{eqn:toeplitz_solution} cannot be evaluated in closed form. If $K=1$, then $C(e^{i\theta}) = q/(q^2-K\beta^2)$ is constant, and we recover \eqref{eqn:k1_solution}. If $K=2$, we can make progress because \eqref{eqn:toeplitz_solution} reduces to the arcsine distribution integral 
\begin{align}
    c_{\tau} = \frac{q}{q^2-2\beta^2} \int_{-\pi}^{\pi} \frac{d\theta}{2\pi} \frac{e^{i\tau\theta}}{1-\alpha \cos(\theta)} 
\end{align}
where
\begin{align}
    \alpha = \frac{2\beta^2 \Gamma_1}{q^2-2\beta^2} . 
\end{align}
Assuming $-1<\alpha < 1$, we can explicitly evaluate the integral to obtain
\begin{align}
    c_{\tau} = \frac{q}{q^2-2\beta^2} \frac{(\sqrt{1-\alpha^2}-1)^{|\tau|}}{\alpha^{|\tau|} \sqrt{1-\alpha^2}} .
\end{align}

In the Gaussian case, upon fixing $q=1$ we are done. We must only verify that the parameters are chosen such that the system is stable, which requires that $|\alpha|<1$ and that $\beta^{2}<1/2$ such that $c_{\tau=0}>0$. This means that we must have 
\begin{align}
    |\Gamma_{1}| < \min\left\{1, \frac{1-2\beta^{2}}{2 \beta^{2} } \right\}
\end{align}
such that $|\alpha|<1$ and such that the weight covariance matrix $\begin{pmatrix} 1 & \Gamma_{1} \\ \Gamma_{1} & 1\end{pmatrix}$ is positive definite. This threshold is consistent with the sufficient condition for stability that we found above. 

In the spherical case, we can use the condition $c_{\tau=0}=1$ to solve for 
\begin{align}
    q = \sqrt{\frac{1+4\beta^2+\sqrt{1+8\beta^2+16\beta^4\Gamma_1^2}}{2}} , 
\end{align}
where we use the condition that $q|_{\beta=0}=1$ to select the appropriate solution. This leads to a value of 
\begin{align} \label{eqn:alpha}
    \alpha = \frac{4\Gamma_1 \beta^2}{1+\sqrt{1+8\beta^2 +16 \beta^4 \Gamma_1^2}}, 
\end{align}
which satisfies $|\alpha|<1$, and the simplified form
\begin{align}
    c_{\tau} = \left( \frac{\sqrt{1-\alpha^2}-1}{\alpha} \right)^{|\tau|}. 
\end{align}
Thus, with a two-step context, we now have exponentially-fast decorrelation with a length scale determined by the correlation strength and temperature.

\subsection{Reverse-engineering weight correlations to achieve a desired stationary correlation}

Instead of choosing $\Gamma_{k}$ and trying to figure out the resulting autocorrelation $c_{\tau}$, we can choose a desired stationary autocorrelation $c_{\tau}$ and try to reverse-engineer the required correlations $\Gamma_{k}$. The latter strategy is more in keeping with the overall goal of this note, which is to design a simple data model with tunable correlations. 

We start by re-writing the $z=e^{i\theta}$-space equation for the stationary autocorrelation in a spherical model as
\begin{align}
    \sum_{k=1-K}^{K-1} \beta^{2} (K-|k|) \Gamma_{k} e^{-ik\theta}  = q^2 - \frac{q}{C(e^{i\theta})}, 
\end{align}
where we assume that $C(e^{i\theta}) \neq 0$. Then, we can use the orthogonality of the Fourier modes to extract
\begin{align} \label{eqn:gamma_solution}
    \Gamma_{k} = \frac{q}{\beta^{2} (K-|k|) } \left( q \delta_{k,0} - \int_{-\pi}^{\pi} \frac{d\theta}{2\pi} e^{i k \theta} \frac{1}{C(e^{i\theta})} \right). 
\end{align}
In this, we have two degrees of freedom: our choice of $q$, and that of $\beta$. These parameters are constrained by self-consistency conditions, including the fact that we must have $\Gamma_{0} > 0$. Moreover, $K$ must be large enough so that all relevant Fourier modes are captured. 

For example, suppose that we want to have autocorrelations that decay exponentially with some rate $\lambda > 0$: 
\begin{align}
    c_{\tau} = e^{-\lambda |\tau|} . 
\end{align}
Then, by evaluating \eqref{eqn:gamma_solution} we find that we should take $K=2$, and 
\begin{align}
    \Gamma_{k} = \frac{q }{2\beta^2} (q - \coth(\lambda)) \delta_{k,0} - \frac{q}{2\beta^{2}} \csch(\lambda) \delta_{|k|, 1}. 
\end{align}
Here, we must clearly choose $q > \coth(\lambda) > 0$, so that $\Gamma_{0}$ is positive. 

But, we can compare this with our previous solution of the self-consistent equation for $K=2$. There we set the temperature such that $\Gamma_{0} = 1$, which means we should take
\begin{align}
    2 \beta^2 = q (q-\coth(\lambda)) \quad \mathrm{and} \quad \Gamma_{1} = -\frac{\csch(\lambda)}{q-\coth(\lambda)}. 
\end{align}
Substituting these values into \eqref{eqn:alpha}, we find that $\alpha = -\sech(\lambda)$ for any $q > \coth(\lambda)$, which leads to an autocorrelation
\begin{align}
    c_{\tau} = \left( \frac{\sqrt{1-\alpha^2}-1}{\alpha} \right)^{|\tau|} = e^{-\lambda |\tau|}. 
\end{align}
Thus, provided that we choose $q$ appropriately, everything is self-consistent. 

\section{Conclusions}

We have analyzed the dynamics of a class of vector spin models generalizing the disordered kinetic Ising model, showing the minimal conditions required to obtain non-trivial correlation structure in the stationary state. Our note is a preliminary step towards an analyses of teacher-student learning in these toy autoregressive sequence models \cite{engel2001statistical}. Moreover, the DMFT equations derived here may be of independent interest in the context of nonequilibrium dynamics \cite{mezard2024structured}, as they generalize those derived for single-step context ($K=1$) models in prior works \cite{aguilera2021unifying,aguilera2023nonequilibrium,bal2023spin}. 

\section*{Acknowledgments}

We thank Blake Bordelon for helpful comments on a previous version of this note. 

\section*{Funding acknowledgments}

At the time of a completion of an initial version of this work in 2023, J.A.Z.-V. and C.P. were supported by NSF Award DMS-2134157 and NSF CAREER Award IIS-2239780. At present, J.A.Z.-V. is supported by the Office of the Director of the National Institutes of Health under Award Number DP5OD037354, and by a Junior Fellowship from the Harvard Society of Fellows. C.P. is presently supported by NSF grant DMS-2134157, NSF CAREER Award IIS-2239780, DARPA grant DIAL-FP-038, a Sloan Research Fellowship, and The William F. Milton Fund from Harvard University. C.P.'s work is been made possible in part by a gift from the Chan Zuckerberg Initiative Foundation to establish the Kempner Institute for the Study of Natural and Artificial Intelligence.

\section*{Author contributions}

JAZ-V conceived the project, performed research, and wrote the paper. CP supervised the project and contributed to review and editing.

\bibliography{refs}

\clearpage 

\appendix 

\section{Ising DMFT}\label{app:ising}

Here we give a detailed derivation of the DMFT in the Ising case. 

\subsection{Averaging over the disorder}

We have the quenched generating function
\begin{align}
    Z[\mathbf{b}] 
    &= \mathbb{E}_{\mathbf{J}} Z[\mathbf{b}; \{\mathbf{J}_{k}\}]
    \\ 
    &= \sum_{\{\mathbf{s}_{t}\}_{t=1}^{T}} \exp\left[\sum_{t=1}^{T} \mathbf{b}_{t} \cdot \mathbf{s}_{t} \right] \mathbb{E}_{\mathbf{J}} \prod_{t=1}^{T} p_{\{\mathbf{J}_{k}\}}(\mathbf{s}_{t}\, |\, \mathbf{s}_{t-1},\ldots \mathbf{s}_{t-K})
    \\ 
    &= \sum_{\{\mathbf{s}_{t}\}_{t=1}^{T}} \exp\left[\sum_{t=1}^{T} \mathbf{b}_{t} \cdot \mathbf{s}_{t} \right] \mathbb{E}_{\mathbf{J}} \prod_{t=1}^{T} \frac{\exp\left[ - \beta \sum_{k=1}^{K} \mathbf{s}_{t}^{\top} \mathbf{J}_{k} \mathbf{s}_{t-k} \right]}{\sum_{\mathbf{u}_{t}} \exp\left[ - \beta \sum_{k=1}^{K} \mathbf{u}_{t}^{\top} \mathbf{J}_{k} \mathbf{s}_{t-k} \right]} .
\end{align}
Under the assumption that the interaction matrices are jointly Gaussian, the local fields
\begin{align}
    \mathbf{h}_{t} = - \beta \sum_{k=1}^{K} \mathbf{J}_{k} \mathbf{s}_{t-k}
\end{align}
are jointly Gaussian under the distribution of the disorder, and have mean zero at all times. With the specific assumption that
\begin{align}
    \mathbb{E}_{\mathbf{J}}[ (J_{k})_{ij} (J_{k'})_{i'j'}] = \frac{1}{N} \delta_{ii'} \delta_{jj'} \Gamma_{k,k'} ,
\end{align}
their covariance is
\begin{align}
    \mathbb{E}_{\mathbf{J}}[h_{t,j} h_{t',j'}] = \delta_{jj'} \beta^{2} \sum_{k,k'=1}^{K} \Gamma_{k,k'} C_{t-k,t'-k'},
\end{align}
where we define the temporal correlation function of the state: 
\begin{align}
    C_{t,t'} = \frac{1}{N} \mathbf{s}_{t}^{\top}\mathbf{s}_{t'} .
\end{align}
As the states are binary, we have 
\begin{align}
    C_{t,t} = 1
\end{align}
at all times.

\subsection{Introducing order parameters}

We now follow the standard procedure of introducing the correlation functions as order parameters. We enforce the definition of $C_{t,t'}$ via Fourier representations of the $\delta$-distribution with Lagrange multipliers $\hat{C}_{t,t'}$, writing
\begin{align}
    1 = \int d\mathbf{C} \int d\hat{\mathbf{C}}\, \exp\left[ - \frac{N}{2} \sum_{t,t'=1-K}^{T} C_{t,t'} \hat{C}_{t,t'} + \frac{1}{2} \sum_{t,t'=1-K}^{T} \hat{C}_{t,t'} \mathbf{s}_{t} \cdot \mathbf{s}_{t'} \right] ,
\end{align}
where the integral over $\mathbf{C}$ is taken over real symmetric matrices with diagonal elements identically equal to one and the integral over $\hat{\mathbf{C}}$ is taken over imaginary symmetric matrices with diagonal elements identically equal to zero. We absorb the factors of $2 \pi$ required to normalize the $\delta$-distribution into the measure $d\hat{\mathbf{C}}$. 

We then have
\begin{align}
    Z 
    &=  \int d\mathbf{C} \int d\hat{\mathbf{C}}\, \exp\left[ - \frac{N}{2} \sum_{t,t'=1-K}^{T} C_{t,t'} \hat{C}_{t,t'}  \right] 
    \nonumber\\&\quad \times  \sum_{\{\mathbf{s}_{t}\}_{t=1}^{T}} \mathbb{E}_{\mathbf{h}} \exp\left[\sum_{t=1}^{T} \mathbf{b}_{t} \cdot \mathbf{s}_{t} + \frac{1}{2} \sum_{t,t'=1-K}^{T} \hat{C}_{t,t'} \mathbf{s}_{t} \cdot \mathbf{s}_{t'}\right] \prod_{t=1}^{T} \frac{\exp[ \mathbf{s}_{t}^{\top} \mathbf{h}_{t} ]}{\sum_{\mathbf{u}_{t}} \exp[ \mathbf{u}_{t}^{\top} \mathbf{h}_{t} ]} .
\end{align}
As the fields $h_{t,j}$ are independent and identically distributed across sites $j$, we have the factorization 
\begin{align}
    & \sum_{\{\mathbf{s}_{t}\}_{t=1}^{T}} \mathbb{E}_{\mathbf{h}} \exp\left[\sum_{t=1}^{T} \mathbf{b}_{t} \cdot \mathbf{s}_{t} + \frac{1}{2} \sum_{t,t'=1-K}^{T} \hat{C}_{t,t'} \mathbf{s}_{t} \cdot \mathbf{s}_{t'}\right] \prod_{t=1}^{T} \frac{\exp[ \mathbf{s}_{t}^{\top} \mathbf{h}_{t} ]}{\sum_{\mathbf{u}_{t}} \exp[ \mathbf{u}_{t}^{\top} \mathbf{h}_{t} ]}
    \\
    &= \sum_{\{\mathbf{s}_{t}\}_{t=1}^{T}} \mathbb{E}_{\mathbf{h}} \exp\left[\sum_{t=1}^{T} \mathbf{b}_{t} \cdot \mathbf{s}_{t} + \frac{1}{2} \sum_{t,t'=1-K}^{T} \hat{C}_{t,t'} \mathbf{s}_{t} \cdot \mathbf{s}_{t'}\right] \prod_{t=1}^{T} \frac{\exp[ \mathbf{s}_{t}^{\top} \mathbf{h}_{t} ]}{\prod_{j=1}^{N} 2 \cosh( h_{t,j}) }
    \\
    &= \prod_{j=1}^{N} z_{j} ,
\end{align}
where we have defined the single-site generating functions
\begin{align}
    z_{j} = \sum_{\{s_{t,j}\}_{t=1}^{T}} \mathbb{E}_{\mathbf{h}_{j}} \exp\left[\sum_{t=1}^{T} b_{t,j} s_{t,j} + \frac{1}{2} \sum_{t,t'=1-K}^{T} \hat{C}_{t,t'} s_{t,j} s_{t',j} \right] \prod_{t=1}^{T} \frac{\exp(s_{t,j} h_{t,j} )}{ 2 \cosh( h_{t,j}) }.
\end{align}
This yields
\begin{align}
    Z =  \int d\mathbf{C} \int d\hat{\mathbf{C}}\, \exp[N S],
\end{align}
where
\begin{align}
    S = - \frac{1}{2} \sum_{t,t'=1-K}^{T} C_{t,t'} \hat{C}_{t,t'} + \frac{1}{N} \sum_{j=1}^{N} \log z_{j}.
\end{align}
At this point, all site-specific dependence is through the initial conditions, which appear in the single-site term multiplying $\hat{C}_{t,t'}$, and through the sources. As a result, the integrals over the order parameters and Lagrange multipliers should be amenable to saddle-point evaluation at large $N$. 

\subsection{The saddle-point equations}

We can now determine the saddle-point equations. From $\partial S/\partial \hat{C}_{t,t'} = 0$, we have the self-consistent equation
\begin{align}
    C_{t,t'} = \frac{1}{N} \sum_{j=1}^{N} \langle s_{t,j} s_{t',j} \rangle_{j}
\end{align}
for any two distinct times $t,t' = 1-K, \ldots, T$. Here, the single-site average $\langle \cdot \rangle_{j}$ is defined via
\begin{align}
    \langle \cdot \rangle_{j} = \frac{1}{z_{j}}  \sum_{\{s_{t,j}\}_{t=1}^{T}} \mathbb{E}_{\mathbf{h}_{j}} (\cdot) \exp\left[\sum_{t=1}^{T} b_{t,j} s_{t,j} + \frac{1}{2} \sum_{t,t'=1-K}^{T} \hat{C}_{t,t'} s_{t,j} s_{t',j} \right] \prod_{t=1}^{T} \frac{\exp(s_{t,j} h_{t,j} )}{ 2 \cosh( h_{t,j}) } \bigg|_{b=0}.
\end{align}
One must be a bit careful when considering boundary terms for which $t$ or $t'$ is less than or equal to zero, as the boundary states are deterministic and fixed by the initial condition. For $t,t' = 1-K, \ldots, 0$, $C_{t,t'}$ is entirely fixed by the boundary condition as
\begin{align}
    C_{t,t'} = \frac{1}{N} \sum_{j=1}^{N} s_{t,j} s_{t',j}.
\end{align}
Now suppose that $t = 1, \ldots, T$ and $t' = 1-K, \ldots, 0$. Then, $s_{t',j}$ is fixed by the initial condition, and we have
\begin{align}
    C_{t,t'} = \frac{1}{N} \sum_{j=1}^{N} \langle s_{t,j} \rangle_{j} s_{t',j}.
\end{align}

Our remaining task is to determine the Lagrange multipliers $\hat{C}_{t,t'}$, recalling that $\hat{C}_{t,t} = 0$ by definition. We first remark that the solution 
\begin{align}
    \hat{C}_{t,t'} = 0
\end{align}
for all $t,t'$ is consistent with the normalization condition $Z=1$ when $b=0$, as we then have
\begin{align}
    z_{j} \bigg|_{b=0} 
    &= \sum_{\{s_{t,j}\}_{t=1}^{T}} \mathbb{E}_{\mathbf{h}_{j}} \prod_{t=1}^{T} \frac{\exp(s_{t,j} h_{t,j} )}{ 2 \cosh( h_{t,j}) }
    \\ 
    &= \mathbb{E}_{\mathbf{h}_{j}} \frac{2 \cosh(h_{t,j})}{2 \cosh(h_{t,j})}
    \\ 
    &= 1,
\end{align}
and $\sum_{t \neq t'} \hat{C}_{t,t'} C_{t,t'} = 0$, which yields $Z|_{b=0} = 1$.

To show this more carefully, we consider the saddle-point equation $\partial S/\partial C_{t,t'} = 0$, which gives
\begin{align}
    \hat{C}_{t,t'} = \frac{1}{N} \sum_{j=1}^{N} \frac{1}{z_{j}} \frac{\partial z_{j}}{\partial C_{t,t'}} \bigg|_{b=0}
\end{align}
for any two distinct times $t,t'$, where we recall that we have constrained $\hat{C}_{t,t'}$ to be symmetric, which compensates for the factor of 1/2. All of the dependence on $C_{t,t'}$ in $z_{j}$ is contained in the covariance of $\mathbf{h}_{j}$, which for any $t,t' = 1, \ldots, T$ is
\begin{align}
    \Sigma_{t,t'} \equiv \mathbb{E}_{\mathbf{J}}[h_{t,j} h_{t',j}] = \beta^{2} \sum_{k,k'=1}^{K} \Gamma_{k,k'} C_{t-k,t'-k'} . 
\end{align}
Then, for any times $q,q' = 1, \ldots, T$ and any two distinct times $t,t' = 1-K, \ldots, T$, we have
\begin{align}
    \frac{\partial \Sigma_{q,q'}}{\partial C_{t,t'}} = \beta^{2} \sum_{k,k'=1}^{K} \Gamma_{k,k'} \delta_{q-k,t} \delta_{q'-k',t'} 
\end{align}
so 
\begin{align}
    \frac{1}{z_{j}}\frac{\partial z_{j}}{\partial C_{t,t'}} \bigg|_{b=0} 
    &= \sum_{q,q' = 1}^{T} \frac{\partial \Sigma_{q,q'}}{\partial C_{t,t'}} \frac{1}{z_{j}}\frac{\partial z_{j}}{\partial \Sigma_{q,q'}} \bigg|_{b=0}
    \\
    &= \beta^{2} \sum_{k,k'=1}^{K} \Gamma_{k,k'} \sum_{q,q' = 1}^{T} \delta_{q-k,t} \delta_{q'-k',t'} \frac{1}{z_{j}}\frac{\partial z_{j}}{\partial \Sigma_{q,q'}} \bigg|_{b=0}
    \\ 
    &= \beta^{2} \sum_{k,k'=1}^{K} \Gamma_{k,k'} \mathbf{1}\{t+k, t'+k' \geq 1\}  \frac{1}{z_{j}}\frac{\partial z_{j}}{\partial \Sigma_{t+k,t'+k'}} \bigg|_{b=0}. 
\end{align}
We can evaluate the required derivatives of expectations using Price's theorem \cite{price1958useful}. However, we must be careful to account for the fact that $t+k$ and $t'+k'$ might coincide. First, for any distinct times $t,t' = 1, \ldots, T$, we have
\begin{align}
    \frac{\partial}{\partial \Sigma_{t,t'}} \mathbb{E}_{\mathbf{h}_{j}}  \prod_{t=1}^{T} \frac{\exp(s_{t,j} h_{t,j} )}{ 2 \cosh( h_{t,j}) }
    &= \mathbb{E}_{\mathbf{h}_{j}} \frac{\partial}{\partial h_{t,j}} \frac{\partial}{\partial h_{t',j}} \prod_{t=1}^{T} \frac{\exp(s_{t,j} h_{t,j} )}{ 2 \cosh( h_{t,j}) }
    \\ 
    &= \mathbb{E}_{\mathbf{h}_{j}} [s_{t,j} - \tanh(h_{t,j})] [s_{t',j} - \tanh(h_{t',j})] \prod_{t''=1}^{T} \frac{\exp(s_{t'',j} h_{t'',j} )}{ 2 \cosh( h_{t'',j}) }.
\end{align}
Second, we have the equal-time derivatives
\begin{align}
    \frac{\partial}{\partial \Sigma_{t,t}} \mathbb{E}_{\mathbf{h}_{j}}  \prod_{t=1}^{T} \frac{\exp(s_{t,j} h_{t,j} )}{ 2 \cosh( h_{t,j}) }
    &= \frac{1}{2} \mathbb{E}_{\mathbf{h}_{j}} \frac{\partial^2}{\partial h_{t,j}^{2}} \prod_{t=1}^{T} \frac{\exp(s_{t,j} h_{t,j} )}{ 2 \cosh( h_{t,j}) }
    \\ 
    &= \frac{1}{2} \mathbb{E}_{\mathbf{h}_{j}} \{[s_{t,j} - \tanh(h_{t,j})]^2 - \sech(h_{t,j})^{2}\} \prod_{t=1}^{T} \frac{\exp(s_{t,j} h_{t,j} )}{ 2 \cosh( h_{t,j}) } . 
\end{align}
Putting these results together and using the definition of the single-site average, we find that for any $t,t' = 1, \ldots, T$
\begin{align}
    \frac{1}{z_{j}}\frac{\partial z_{j}}{\partial \Sigma_{t,t'}} \bigg|_{b=0}
    &= (1-\delta_{t,t'}) \bigg\langle [s_{t,j} - \tanh(h_{t,j})] [s_{t',j} - \tanh(h_{t',j})] \bigg\rangle_{j} \nonumber\\&\quad + \frac{1}{2} \delta_{t,t'} \bigg\langle [s_{t,j} - \tanh(h_{t,j})]^2 - \sech(h_{t,j})^{2} \bigg\rangle_{j} , 
\end{align}
in terms of which we have
\begin{align}
    \hat{C}_{t,t'} 
    &= \beta^{2} \sum_{k,k'=1}^{K} \Gamma_{k,k'} \mathbf{1}\{t+k, t'+k' \geq 1\}  \frac{1}{N} \sum_{j=1}^{N} \frac{1}{z_{j}}\frac{\partial z_{j}}{\partial \Sigma_{t+k,t'+k'}} \bigg|_{b=0}
\end{align}
for any two distinct times $t,t'$. We emphasize that thanks to the constraint on the sum over $k$ and $k'$, the right-hand-side of this equation does not depend directly on the initial condition; all of the fields that appear are at times greater than or equal to 1.

Then, to show that the solution $\hat{C}_{t,t'} = 0$ is self-consistent, we would like to show that the expectations listed above vanish under that assumption. To do so, we will simplify the single-site generating function
\begin{align}
    z_{j} = \sum_{\{s_{t,j}\}_{t=1}^{T}} \mathbb{E}_{\mathbf{h}_{j}} \exp\left[\sum_{t=1}^{T} b_{t,j} s_{t,j} + \frac{1}{2} \sum_{t,t'=1-K}^{T} \hat{C}_{t,t'} s_{t,j} s_{t',j} \right] \prod_{t=1}^{T} \frac{\exp(s_{t,j} h_{t,j} )}{ 2 \cosh( h_{t,j}) }
\end{align}
and the resulting single-site averages using a Hubbard-Stratonovich transformation. Noting that $\hat{C}_{t,t} = 0$ by construction, we multiply and divide by 
\begin{align}
    \exp\left(\frac{1}{2} (T+K) \rho \right)
\end{align}
for a positive real regulating parameter $\rho$, which gives
\begin{align}
    z_{j} = e^{-(T+K)\rho/2} \sum_{\{s_{t,j}\}_{t=1}^{T}} \mathbb{E}_{\mathbf{h}_{j}} \exp\left[\sum_{t=1}^{T} b_{t,j} s_{t,j} + \frac{1}{2} \sum_{t,t'=1-K}^{T} (\rho\delta_{t,t'} + \hat{C}_{t,t'}) s_{t,j} s_{t',j} \right] \prod_{t=1}^{T} \frac{\exp(s_{t,j} h_{t,j} )}{ 2 \cosh( h_{t,j}) } .
\end{align}
Then, letting $u_{t,j}$ be a mean-zero Gaussian field with covariance
\begin{align}
    \mathbb{E}_{\mathbf{u}_{j}}[u_{t,j} u_{t',j}] = \rho\delta_{t,t'} + \hat{C}_{t,t'},
\end{align}
we have
\begin{align}
    z_{j} &= e^{-(T+K)\rho/2} \mathbb{E}_{\mathbf{h}_{j},\mathbf{u}_{j}} \sum_{\{s_{t,j}\}_{t=1}^{T}} \exp\left[\sum_{t=1}^{T} b_{t,j} s_{t,j} + \sum_{t=1-K}^{T} u_{t,j} s_{t,j}  \right] \prod_{t=1}^{T} \frac{\exp(s_{t,j} h_{t,j} )}{ 2 \cosh( h_{t,j}) } 
    \\ 
    &= e^{-(T+K)\rho/2} \mathbb{E}_{\mathbf{h}_{j},\mathbf{u}_{j}} \exp\left[ \sum_{t=1-K}^{0} u_{t,j} s_{t,j} \right] \prod_{t=1}^{T} \frac{\sum_{s_{t,j}} \exp(s_{t,j} b_{t,j} + s_{t,j} h_{t,j} + s_{t,j} u_{t,j})}{2\cosh(h_{t,j})}
    \\ 
    &= e^{-(T+K)\rho/2} \mathbb{E}_{\mathbf{h}_{j},\mathbf{u}_{j}} \exp\left[ \sum_{t=1-K}^{0} u_{t,j} s_{t,j} \right] \prod_{t=1}^{T} \frac{\cosh(b_{t,j} + h_{t,j} + u_{t,j})}{\cosh(h_{t,j})} .
\end{align}
Then, for any two distinct positive times $t,t' = 1, \ldots, T$, we have
\begin{align}
    \langle s_{t,j} \rangle_{j} 
    &= \frac{1}{z_{j}} \frac{\partial z_{j}}{\partial b_{t,j}} \bigg|_{b=0}
    \\ 
    &=  \frac{1}{z_{j}} e^{-(T+K)\rho/2} \mathbb{E}_{\mathbf{v}_{j},\mathbf{u}_{j}}\Bigg\{ \exp\left[\sum_{t=1-K}^{0} s_{t,j} u_{t,j} \right] \left[ \prod_{t' \neq t} \frac{\cosh(b_{t',j} + h_{t',j} + u_{t',j})}{\cosh(h_{t',j})} \right] \nonumber\\&\qquad\qquad\qquad\qquad\qquad\times  \frac{\sinh(b_{t,j} + h_{t,j} + u_{t,j})}{\cosh(h_{t,j})} \Bigg\} \bigg|_{b=0}
    \\ 
    &= \langle \tanh( h_{t,j} + u_{t,j}) \rangle_{j} ,
\end{align}
\begin{align}
    \langle s_{t,j} s_{t',j} \rangle_{j} 
    &= \frac{1}{z_{j}} \frac{\partial^2 z_{j}}{\partial b_{t,j} \partial b_{t',j}} \bigg|_{b=0}
    \\ 
    &=  \frac{1}{z_{j}} e^{-(T+K)\rho/2} \mathbb{E}_{\mathbf{v}_{j},\mathbf{u}_{j}}\Bigg\{ \exp\left[\sum_{t=1-K}^{0} s_{t,j} u_{t,j} \right] \left[ \prod_{t'' \neq t,t'} \frac{\cosh(b_{t'',j} + h_{t'',j} + u_{t'',j})}{\cosh(h_{t'',j})} \right] \nonumber\\&\qquad\qquad\qquad\qquad\qquad\times \frac{\sinh(b_{t,j} + h_{t,j} + u_{t,j})}{\cosh(h_{t,j})} \frac{\sinh(b_{t',j} + h_{t',j} + u_{t',j})}{\cosh(h_{t',j})}  \Bigg\} \bigg|_{b=0}
    \\ 
    &= \langle \tanh( h_{t,j} + u_{t,j}) \tanh( h_{t',j} + u_{t',j}) \rangle_{j} ,
\end{align}
and
\begin{align}
    \langle s_{t,j} \tanh(h_{t',j}) \rangle_{j} 
    &= \frac{1}{z_{j}} e^{-(T+K)\rho/2} \mathbb{E}_{\mathbf{v}_{j},\mathbf{u}_{j}}\Bigg\{ \exp\left[\sum_{t=1-K}^{0} s_{t,j} u_{t,j} \right] \left[ \prod_{t'' \neq t} \frac{\cosh(b_{t'',j} + h_{t'',j} + u_{t'',j})}{\cosh(h_{t'',j})} \right] \nonumber\\&\qquad\qquad\qquad\qquad\qquad\times \frac{\sinh(b_{t,j} + h_{t,j} + u_{t,j})}{\cosh(h_{t,j})} \tanh(h_{t',j}) \Bigg\} \bigg|_{b=0}
    \\ 
    &= \langle \tanh( h_{t,j} + u_{t,j}) \tanh( h_{t',j} ) \rangle_{j} . 
\end{align}
Examining these results, if $\hat{C}_{t,t'} = 0$, then the correlations between different $u_{t,j}$ vanish, and we can evaluate all of the expectations over the then independent variables $u_{t,j} \sim \mathcal{N}(0,\rho)$ using the identity
\begin{align}
    \mathbb{E}_{x \sim \mathcal{N}(0,\rho)} \sinh(x+y) = e^{\rho/2} \sinh(y) .
\end{align}
In particular, all of the dependence on the regulator $\rho$ drops out---as should happen as it is arbitrary---and we can see that we will then have
\begin{align}
    \langle s_{t,j} \rangle_{j} \bigg|_{\hat{C}_{t,t'} = 0} = \langle \tanh(h_{t,j}) \rangle_{j} ,
\end{align}
\begin{align}
    \langle s_{t,j} s_{t',j} \rangle_{j} \bigg|_{\hat{C}_{t,t'} = 0} = \langle \tanh(h_{t,j}) \tanh(h_{t',j}) \rangle_{j},
\end{align}
and
\begin{align}
    \langle s_{t,j} \tanh(h_{t',j}) \rangle_{j} \bigg|_{\hat{C}_{t,t'} = 0}
    &= \langle \tanh( h_{t,j} ) \tanh( h_{t',j} ) \rangle_{j} 
\end{align}
for any two distinct times $t,t' = 1, \ldots, T$.

Therefore, for any two distinct times $t,t' = 1,\ldots T$, we have
\begin{align}
    \frac{1}{z_{j}}\frac{\partial z_{j}}{\partial \Sigma_{t,t'}} \bigg|_{b=0}\bigg|_{\hat{C}_{t,t'} = 0} = \bigg\langle [s_{t,j} - \tanh(h_{t,j})] [s_{t',j} - \tanh(h_{t',j})] \bigg\rangle_{j} \bigg|_{\hat{C}_{t,t'} = 0} 
    &= 0 . 
\end{align}

We now turn our attention to the $t = t'$ term 
\begin{align}
    \frac{1}{z_{j}}\frac{\partial z_{j}}{\partial \Sigma_{t,t}} \bigg|_{b=0}
    &= \frac{1}{2} \bigg\langle [s_{t,j} - \tanh(h_{t,j})]^2 - \sech(h_{t,j})^{2} \bigg\rangle_{j} .
\end{align} 
As $s_{t,j}$ is binary, we have
\begin{align}
    \bigg\langle [s_{t,j} - \tanh(h_{t,j})]^2 - \sech(h_{t,j})^{2} \bigg\rangle_{j} 
    &= 1 - 2 \langle s_{t,j} \tanh(h_{t,j}) \rangle_{j} + \langle \tanh(h_{t,j})^2 \rangle_{j} - \langle \sech(h_{t,j})^2 \rangle_{j}  .
\end{align}
Using the result above, we may write
\begin{align}
    \langle s_{t,j} \tanh(h_{t,j}) \rangle_{j} 
    &= \frac{1}{z_{j}} e^{-(T+K)\rho/2} \mathbb{E}_{\mathbf{v}_{j},\mathbf{u}_{j}}\Bigg\{ \exp\left[\sum_{t=1-K}^{0} s_{t,j} u_{t,j} \right] \left[ \prod_{t'' \neq t} \frac{\cosh(b_{t'',j} + h_{t'',j} + u_{t'',j})}{\cosh(h_{t'',j})} \right] \nonumber\\&\qquad\qquad\qquad\qquad\qquad\times \frac{\sinh(b_{t,j} + h_{t,j} + u_{t,j})}{\cosh(h_{t,j})} \tanh(h_{t,j}) \Bigg\} \bigg|_{b=0}
    \\ 
    &= \langle \tanh( h_{t,j} + u_{t,j}) \tanh( h_{t,j} ) \rangle_{j} . 
\end{align}
From this, and following the logic above for how to evalaute the expectation over $u_{t,j}$ when $\hat{C}_{t,t'} = 0$, we have
\begin{align}
    \bigg\langle [s_{t,j} - \tanh(h_{t,j})]^2 - \sech(h_{t,j})^{2} \bigg\rangle_{j} \bigg|_{\hat{C}_{t,t'} = 0} 
    &= 1 - 2 \langle \tanh(h_{t,j})^2 \rangle_{j} + \langle \tanh(h_{t,j})^2 \rangle_{j} - \langle \sech(h_{t,j})^2 \rangle_{j} 
    \\ 
    &= \bigg\langle 1 - [\tanh(h_{t,j})^2 + \sech(h_{t,j})^2] \bigg\rangle_{j} 
    \\ 
    &= 0
\end{align}
thanks to the identity $\tanh(h)^2 + \sech(h)^2 = 1$. 

Therefore, we conclude that for any two times $t,t' = 1, \ldots, T$, whether equal or not, we have
\begin{align}
    \frac{1}{z_{j}}\frac{\partial z_{j}}{\partial \Sigma_{t,t'}} \bigg|_{b=0} \bigg|_{\hat{C}_{t,t'} = 0} = 0. 
\end{align}
Thus, as the saddle-point equation for $\hat{C}_{t,t'}$ is
\begin{align}
    \hat{C}_{t,t'} 
    &= \beta^{2} \sum_{k,k'=1}^{K} \Gamma_{k,k'} \mathbf{1}\{t+k, t'+k' \geq 1\}  \frac{1}{N} \sum_{j=1}^{N} \frac{1}{z_{j}}\frac{\partial z_{j}}{\partial \Sigma_{t+k,t'+k'}} \bigg|_{b=0},
\end{align}
we find that the solution
\begin{align}
    \hat{C}_{t,t'} = 0
\end{align}
is self-consistent, as suggested by the normalization of the generating function.

Moreover, with $\hat{C}_{t,t'} = 0$ the single-site generating function simplifies to  
\begin{align}
    z_{j} 
    &= \sum_{\{s_{t,j}\}_{t=1}^{T}} \mathbb{E}_{\mathbf{h}_{j}} \exp\left[\sum_{t=1}^{T} b_{t,j} s_{t,j} \right] \prod_{t=1}^{T} \frac{\exp(s_{t,j} h_{t,j} )}{ 2 \cosh( h_{t,j}) }
    \\ 
    &= \mathbb{E}_{\mathbf{h}_{j}} \prod_{t=1}^{T} \frac{\cosh(b_{t,j} + h_{t,j})}{\cosh(h_{t,j})}
\end{align}
and the single-site averages simplify accordingly. In particular, for any two distinct times $t,t' = 1, \ldots, T$, we now have
\begin{align}
    C_{t,t'} 
    &= \frac{1}{N} \sum_{j=1}^{N} \langle s_{t,j} s_{t',j} \rangle_{j}
    \\ 
    &= \mathbb{E}_{\mathbf{h}}[\tanh(h_{t}) \tanh(h_{t'})],
\end{align}
as the single-site average is now manifestly site-independent. As noted previously, for $t,t' = 1-K, \ldots, 0$, $C_{t,t'}$ is entirely fixed by the boundary condition as
\begin{align}
    C_{t,t'} = \frac{1}{N} \sum_{j=1}^{N} s_{t,j} s_{t',j}.
\end{align}
Finally, for mixed boundary-bulk correlators with $t = 1, \ldots, T$ and $t' = 1-K, \ldots, 0$, we have
\begin{align}
    C_{t,t'} 
    &= \frac{1}{N} \sum_{j=1}^{N} \langle s_{t,j} \rangle_{j} s_{t',j} 
    \\ 
    &=  \frac{1}{N} \sum_{j=1}^{N} s_{t',j} \mathbb{E}_{\mathbf{v}}[\tanh(v_{t})] 
    \\ 
    &= 0.
\end{align}

\subsection{Summarizing the DMFT}

In summary, we have derived a DMFT for the temporal correlation of the state, defined by the self-consistent equation
\begin{align}
    C_{t,t'} = \mathbb{E}_{\mathbf{h}}[\tanh(h_{t})\tanh(h_{t'})]
\end{align}
for distinct times $t,t'=1,\ldots,T$ and $C_{t,t} = 1$, where $h_{t}$ is a zero-mean Gaussian with covariance 
\begin{align}
    \mathbb{E}_{\mathbf{h}}[h_{t} h_{t'}] = \beta^{2} \sum_{k,k'=1}^{K} \Gamma_{k,k'} C_{t-k,t'-k'},
\end{align}
For $t,t' = 1-K, \ldots, 0$, $C_{t,t'}$ is entirely fixed by the initial condition as
\begin{align}
    C_{t,t'} = \frac{1}{N} \sum_{j=1}^{N} s_{t,j} s_{t',j},
\end{align}
while $C_{t,t'} = 0$ if $t = 1, \ldots, T$ and $t' = 1-K, \ldots, 0$.

\section{Gaussian DMFT}\label{app:gaussian}

We now consider the Gaussian model. Here, we fix a prior over states
\begin{align}
    \mathbf{s}_{t} \sim \mathcal{N}(\mathbf{0},\mathbf{I}_{N}),
\end{align}
which in the large-$N$ limit will give a concentrating norm $\Vert \mathbf{s}_{t} \Vert^2 \sim N$. Writing the local field at time $t$ as
\begin{align}
    \mathbf{h}_{t} = - \beta \sum_{k=1}^{K} \mathbf{J}_{k} \mathbf{s}_{t-k}
\end{align}
the normalization factor at that time is then
\begin{align}
    \mathbb{E}_{\mathbf{u}_{t}\sim \mathcal{N}(\mathbf{0},\mathbf{I}_{N})} \exp\left[ - \beta \sum_{k=1}^{K} \mathbf{u}_{t}^{\top} \mathbf{J}_{k} \mathbf{s}_{t-k} \right] 
    &= \mathbb{E}_{\mathbf{u}_{t}\sim \mathcal{N}(\mathbf{0},\mathbf{I}_{N})} \exp\left[ \mathbf{u}_{t}^{\top} \mathbf{h}_{t} \right]
    \\ 
    &= \exp\left(\frac{1}{2} \Vert \mathbf{h}_{t} \Vert^2 \right),
\end{align}
which of course factorizes over sites. We can then write the single-step transition density as 
\begin{align}
    p_{\{\mathbf{J}_{k}\}}(\mathbf{s}_{t} \,|\, \mathbf{s}_{t-1},\ldots,\mathbf{s}_{t-K}) = \exp\left( \mathbf{s}_{t}^{\top} \mathbf{h}_{t} - \frac{1}{2} \Vert \mathbf{h}_{t} \Vert^2 \right),
\end{align}
hence the density with respect to Lebesgue measure simply becomes
\begin{align}
    p_{\{\mathbf{J}_{k}\}}(\mathbf{s}_{t} \,|\, \mathbf{s}_{t-1},\ldots,\mathbf{s}_{t-K}) \frac{d\sigma_{N}}{d\mathbf{s}}(\mathbf{s}_{t}) = \exp\left( - \frac{1}{2} \Vert \mathbf{s}_{t} - \mathbf{h}_{t} \Vert^{2} \right).
\end{align}

\subsection{Averaging over the disorder}

With this setup, the quenched generating function is 
\begin{align}
    Z[\mathbf{b}] 
    &= \mathbb{E}_{\{\mathbf{s}_{t}\sim \mathcal{N}(\mathbf{0},\mathbf{I}_{N})\}} \exp\left[\sum_{t=1}^{T} \mathbf{b}_{t} \cdot \mathbf{s}_{t} \right] \mathbb{E}_{\mathbf{J}} \exp\left[ \sum_{t=1}^{T} \mathbf{s}_{t}^{\top} \mathbf{h}_{t} - \frac{1}{2} \sum_{t=1}^{T} \Vert \mathbf{h}_{t} \Vert^2 \right]. 
\end{align}
As in the Ising case, with the specific assumption that the interaction matrices are zero-mean Gaussian with covariance
\begin{align}
    \mathbb{E}_{\mathbf{J}}[ (J_{k})_{ij} (J_{k'})_{i'j'}] = \frac{1}{N} \delta_{ii'} \delta_{jj'} \Gamma_{k,k'} ,
\end{align}
the local fields are zero-mean Gaussian with covariance
\begin{align}
    \mathbb{E}_{\mathbf{J}}[h_{t,j} h_{t',j'}] = \delta_{jj'} \beta^{2} \sum_{k,k'=1}^{K} \Gamma_{k,k'} C_{t-k,t'-k'},
\end{align}
where we define the temporal correlation function of the state: 
\begin{align}
    C_{t,t'} = \frac{1}{N} \mathbf{s}_{t}^{\top}\mathbf{s}_{t'} .
\end{align}
However, in this case the norm of the state $C_{t,t'}$ is not fixed to unity. 

\subsection{Introducing order parameters}

Again, we enforce the definition of $C_{t,t'}$ via Fourier representations of the $\delta$-distribution with Lagrange multipliers $\hat{C}_{t,t'}$, writing
\begin{align}
    1 = \int d\mathbf{C} \int d\hat{\mathbf{C}}\, \exp\left[ - \frac{N}{2} \sum_{t,t'=1-K}^{T} C_{t,t'} \hat{C}_{t,t'} + \frac{1}{2} \sum_{t,t'=1-K}^{T} \hat{C}_{t,t'} \mathbf{s}_{t} \cdot \mathbf{s}_{t'} \right] ,
\end{align}
where in this case we include the diagonal elements of $C_{t,t'}$ and $\hat{C}_{t,t'}$. Then, everything factors over sites, yielding
\begin{align}
    Z =  \int d\mathbf{C} \int d\hat{\mathbf{C}}\, \exp[N S],
\end{align}
where
\begin{align}
    S = - \frac{1}{2} \sum_{t,t'=1-K}^{T} C_{t,t'} \hat{C}_{t,t'} + \frac{1}{N} \sum_{j=1}^{N} \log z_{j}
\end{align}
for single-site generating functions
\begin{align}
    z_{j} = \mathbb{E}_{\{s_{t,j} \sim \mathcal{N}(0,1)\}} \mathbb{E}_{\mathbf{h}_{j}} \exp\left[\sum_{t=1}^{T} b_{t,j} s_{t,j} + \frac{1}{2} \sum_{t,t'=1-K}^{T} \hat{C}_{t,t'} s_{t,j} s_{t',j} + \sum_{t=1}^{T} s_{t,j} h_{t,j} - \frac{1}{2} \sum_{t=1}^{T} h_{t,j}^{2} \right].
\end{align}
\subsection{The saddle-point equations}

We can now determine the saddle-point equations. From $\partial S/\partial \hat{C}_{t,t'} = 0$, we have the self-consistent equation
\begin{align}
    C_{t,t'} = \frac{1}{N} \sum_{j=1}^{N} \langle s_{t,j} s_{t',j} \rangle_{j}
\end{align}
for any two (non necessarily distinct) times $t,t' = 1-K, \ldots, T$. Here, the self-consistent average is defined via 
\begin{align}
    \langle \cdot \rangle_{j} = \frac{1}{z_{j}} \mathbb{E}_{\{s_{t,j} \sim \mathcal{N}(0,1)\}} \mathbb{E}_{\mathbf{h}_{j}} (\cdot) \exp\left[\sum_{t=1}^{T} b_{t,j} s_{t,j} + \frac{1}{2} \sum_{t,t'=1-K}^{T} \hat{C}_{t,t'} s_{t,j} s_{t',j} + \sum_{t=1}^{T} s_{t,j} h_{t,j} - \frac{1}{2} \sum_{t=1}^{T} h_{t,j}^{2} \right] \bigg|_{b=0}.
\end{align}
Again, one must be careful when considering boundary terms for which $t$ or $t'$ is less than or equal to zero, as the boundary states are fixed by the initial condition. For $t,t' = 1-K, \ldots, 0$, $C_{t,t'}$ is entirely fixed by the boundary condition as
\begin{align}
    C_{t,t'} = \frac{1}{N} \sum_{j=1}^{N} s_{t,j} s_{t',j}.
\end{align}
Now suppose that $t = 1, \ldots, T$ and $t' = 1-K, \ldots, 0$. Then, $s_{t',j}$ is fixed by the initial condition, and we have
\begin{align}
    C_{t,t'} = \frac{1}{N} \sum_{j=1}^{N} \langle s_{t,j} \rangle_{j} s_{t',j}.
\end{align}

We now consider $\hat{C}_{t,t'}$. As in the Ising case, it is easily seen that having $\hat{C}_{t,t'} = 0$ for all times is consistent with the normalization, as we then have
\begin{align}
    z_{j} \bigg|_{b=0} 
    &= \mathbb{E}_{\{s_{t,j} \sim \mathcal{N}(0,1)\}} \mathbb{E}_{\mathbf{h}_{j}} \exp\left[  \sum_{t=1}^{T} s_{t,j} h_{t,j} - \frac{1}{2} \sum_{t=1}^{T} h_{t,j}^{2} \right]
    \\ 
    &= \mathbb{E}_{\mathbf{h}_{j}} \exp\left[ \frac{1}{2} \sum_{t=1}^{T} h_{t,j}^{2}  - \frac{1}{2} \sum_{t=1}^{T} h_{t,j}^{2} \right]
    \\ 
    &= 1.
\end{align}
As we did in the Ising case, we can show this in greater detail by considering the saddle-point equation $\partial S/\partial C_{t,t'} = 0$, which yields
\begin{align}
    \frac{1}{2} \hat{C}_{t,t'} = \frac{1}{N} \sum_{j=1}^{N} \frac{1}{z_{j}} \frac{\partial z_{j}}{\partial C_{t,t}} \bigg|_{b=0}
\end{align}
for $t = t'$ and
\begin{align}
    \hat{C}_{t,t'} = \frac{1}{N} \sum_{j=1}^{N} \frac{1}{z_{j}} \frac{\partial z_{j}}{\partial C_{t,t'}} \bigg|_{b=0}
\end{align}
for $t \neq t'$, where the factor of 1/2 is not present in the off-diagonal terms due to symmetry. Again, writing 
\begin{align}
    \Sigma_{t,t'} \equiv \mathbb{E}_{\mathbf{J}}[h_{t,j} h_{t',j}] = \beta^{2} \sum_{k,k'=1}^{K} \Gamma_{k,k'} C_{t-k,t'-k'} , 
\end{align}
we have
\begin{align}
    \frac{1}{z_{j}}\frac{\partial z_{j}}{\partial C_{t,t'}} \bigg|_{b=0} = \beta^{2} \sum_{k,k'=1}^{K} \Gamma_{k,k'} \mathbf{1}\{t+k, t'+k' \geq 1\}  \frac{1}{z_{j}}\frac{\partial z_{j}}{\partial \Sigma_{t+k,t'+k'}} \bigg|_{b=0}. 
\end{align}
Applying Price's theorem \cite{price1958useful}, we have for any two distinct times $t,t'$
\begin{align}
    \frac{\partial}{\partial \Sigma_{t,t'}} \mathbb{E}_{\mathbf{h}_{j}} \exp\left[ \sum_{t=1}^{T} s_{t,j} h_{t,j} - \frac{1}{2} \sum_{t=1}^{T} h_{t,j}^2 \right] 
    &= \mathbb{E}_{\mathbf{h}_{j}} \frac{\partial}{\partial h_{t,j}} \frac{\partial}{\partial h_{t',j}} \exp\left[ \sum_{t=1}^{T} s_{t,j} h_{t,j} - \frac{1}{2} \sum_{t=1}^{T} h_{t,j}^2 \right]
    \\ 
    &= \mathbb{E}_{\mathbf{h}_{j}} (s_{t,j} - h_{t,j}) (s_{t',j} - h_{t',j}) \exp\left[ \sum_{t=1}^{T} s_{t,j} h_{t,j} - \frac{1}{2} \sum_{t=1}^{T} h_{t,j}^2 \right], 
\end{align}
while the equal-time derivatives are
\begin{align}
    \frac{\partial}{\partial \Sigma_{t,t}} \mathbb{E}_{\mathbf{h}_{j}} \exp\left[ \sum_{t=1}^{T} s_{t,j} h_{t,j} - \frac{1}{2} \sum_{t=1}^{T} h_{t,j}^2 \right] 
    &= \frac{1}{2} \mathbb{E}_{\mathbf{h}_{j}} \frac{\partial^2}{\partial h_{t,j}^2} \exp\left[ \sum_{t=1}^{T} s_{t,j} h_{t,j} - \frac{1}{2} \sum_{t=1}^{T} h_{t,j}^2 \right]
    \\ 
    &= \mathbb{E}_{\mathbf{h}_{j}} [(s_{t,j} - h_{t,j})^2 - 1] \exp\left[ \sum_{t=1}^{T} s_{t,j} h_{t,j} - \frac{1}{2} \sum_{t=1}^{T} h_{t,j}^2 \right] . 
\end{align}
Then, using the definition of the single-site average, we have 
\begin{align}
    \frac{1}{z_{j}} \frac{\partial z_{j}}{\partial \Sigma_{t,t'}} \bigg|_{b=0} = \bigg\langle (s_{t,j} - h_{t,j}) (s_{t',j} - h_{t',j}) \bigg\rangle_{j}
\end{align}
for any two distinct times $t,t' = 1, \ldots, T$, while
\begin{align}
    \frac{1}{z_{j}} \frac{\partial z_{j}}{\partial \Sigma_{t,t}} \bigg|_{b=0} = \bigg\langle [(s_{t,j} - h_{t,j})^2 - 1] \bigg\rangle_{j}
\end{align}
for any time $t = 1, \ldots, T$. Thus, as the saddle-point equation gives $\hat{C}_{t,t'}$ as a convex combination of these derivatives, to show that it is self-consistent to have $\hat{C}_{t,t'} = 0$ we would like to show that these averages vanish under that assumption. Unlike in the Ising case, this is easily done thanks to the fact that $s_{t,j}$ and $h_{t,j}$ are jointly Gaussian even for non-vanishing $\hat{C}_{t,t'}$. Recalling the definition of the single-site average as
\begin{align}
    \langle \cdot \rangle_{j} = \frac{1}{z_{j}|_{b=0}} \mathbb{E}_{\{s_{t,j} \sim \mathcal{N}(0,1)\}} \mathbb{E}_{\mathbf{h}_{j}} (\cdot) \exp\left[+ \frac{1}{2} \sum_{t,t'=1-K}^{T} \hat{C}_{t,t'} s_{t,j} s_{t',j} + \sum_{t=1}^{T} s_{t,j} h_{t,j} - \frac{1}{2} \sum_{t=1}^{T} h_{t,j}^{2} \right] ,
\end{align}
we expand the term involving $\hat{C}_{t,t'}$ as 
\begin{align}
    \frac{1}{2} \sum_{t,t'=1-K}^{T} \hat{C}_{t,t'} s_{t,j} s_{t',j} = \frac{1}{2} \sum_{t,t'=1}^{T} \hat{C}_{t,t'} s_{t,j} s_{t',j} + \sum_{t=1}^{T} \sum_{t'=1-K}^{0} \hat{C}_{t,t'} s_{t,j} s_{t',j} + \frac{1}{2} \sum_{t,t'=1-K}^{0} \hat{C}_{t,t'} s_{t,j} s_{t',j} .
\end{align}
The term that depends only on the boundary conditions will cancel in the ratio, which leaves
\begin{align}
    \langle \cdot \rangle_{j} 
    &= \frac{1}{z_{j}|_{b=0}} \int d\mathbf{s}\,\int d\mathbf{h}\, \exp\left[ - \frac{1}{2} \begin{pmatrix} \mathbf{s} \\ \mathbf{h} \end{pmatrix}^{\top} \mathbf{J} \begin{pmatrix} \mathbf{s} \\ \mathbf{h} \end{pmatrix} + \begin{pmatrix} \mathbf{s} \\ \mathbf{h} \end{pmatrix}^{\top} \begin{pmatrix} \mathbf{c} \\ \mathbf{0} \end{pmatrix} \right] ,
\end{align}
where we define the matrix
\begin{align}
    \mathbf{J}
    = \begin{pmatrix}
        \mathbf{I}_{T} - \hat{\mathbf{C}} & -\mathbf{I}_{T} \\ 
        - \mathbf{I}_{T} & \mathbf{I}_{T} + \mathbf{\Sigma}^{-1}
    \end{pmatrix}
\end{align}
and the vector
\begin{align}
    c_{t} = \sum_{t'=1-K}^{0} \hat{C}_{t,t'} s_{t',j} .
\end{align}
Completing the square, this implies that the single-site average is computed for
\begin{align}
    \begin{pmatrix} \mathbf{s} \\ \mathbf{h} \end{pmatrix} \sim \mathcal{N}\left( \mathbf{J}^{-1} \begin{pmatrix} \mathbf{c} \\ \mathbf{0} \end{pmatrix} , \mathbf{J}^{-1} \right) .
\end{align}
Our task is now to invert the block matrix $\mathbf{J}$. Noting that the push-through identity implies that
\begin{align}
    (\mathbf{I}_{T} + \mathbf{\Sigma}^{-1})^{-1} = \mathbf{I}_{T} - (\mathbf{I}_{T} + \mathbf{\Sigma})^{-1},
\end{align}
the formula for $2 \times 2$ block matrix inversion implies that
\begin{align}
    \mathbf{J}^{-1} &= \begin{pmatrix}
        \mathbf{I}_{T} - \hat{\mathbf{C}} & -\mathbf{I}_{T} \\ 
        - \mathbf{I}_{T} & \mathbf{I}_{T} + \mathbf{\Sigma}^{-1}
    \end{pmatrix}^{-1}
    \\
    &= 
    \begin{pmatrix}
        [ (\mathbf{I}_{T} + \mathbf{\Sigma})^{-1} - \hat{\mathbf{C}}]^{-1} &  \mathbf{B}
        \\ 
         \mathbf{B}^{\top} & [\mathbf{I}_{T} - (\mathbf{I}_{T} + \mathbf{\Sigma})^{-1}] ( \mathbf{I}_{T} +  \mathbf{B} )
    \end{pmatrix}
\end{align}
where we write
\begin{align}
    \mathbf{B} &= [ (\mathbf{I}_{T} + \mathbf{\Sigma})^{-1} - \hat{\mathbf{C}} ]^{-1} [\mathbf{I}_{T} - (\mathbf{I}_{T} + \mathbf{\Sigma})^{-1}]
    \\
    &= [\mathbf{I}_{T} - (\mathbf{I}_{T} + \mathbf{\Sigma}) \hat{\mathbf{C}}]^{-1} \mathbf{\Sigma} .
\end{align}
Then, we have
\begin{align}
    \mathbf{J}^{-1} \begin{pmatrix} \mathbf{c} \\ \mathbf{0} \end{pmatrix} = \begin{pmatrix}
        [ (\mathbf{I}_{T} + \mathbf{\Sigma})^{-1} - \hat{\mathbf{C}}]^{-1} \mathbf{c} \\ \mathbf{B}^{\top} \mathbf{c}
    \end{pmatrix} ,
\end{align}
hence
\begin{align}
    \langle s_{t,j} - h_{t,j} \rangle_{j} 
    &= [ (\mathbf{I}_{T} + \mathbf{\Sigma})^{-1} - \hat{\mathbf{C}}]^{-1} \mathbf{c} - \mathbf{B}^{\top} \mathbf{c}
    \\ 
    &= [\mathbf{I}_{T} - \hat{\mathbf{C}} (\mathbf{I}_{T} + \mathbf{\Sigma}) ]^{-1} \mathbf{c} .
\end{align}
Then, noting that
\begin{align}
    \mathbf{J}^{-1} \bigg|_{\hat{C} = 0} = 
    \begin{pmatrix}
        \mathbf{I}_{T} + \mathbf{\Sigma} &  \mathbf{\Sigma}
        \\ 
         \mathbf{\Sigma} & \mathbf{\Sigma},
    \end{pmatrix}
\end{align}
we can then see that, if $\hat{C}_{t,t'} = 0$ for all $t,t'$, we have the single-site averages
\begin{align}
    \langle s_{t,j} - h_{t,j} \rangle_{j} &= 0
    \\ 
    \langle s_{t,j} s_{t',j} \rangle_{j} &= \delta_{t,t'} + \Sigma_{t,t'} 
    \\ 
    \langle s_{t,j} h_{t',j} \rangle_{j} &= \Sigma_{t,t'}
    \\ 
    \langle h_{t,j} h_{t',j} \rangle_{j} &= \Sigma_{t,t'} .
\end{align}
From this, it is easy to see that, for $t \neq t'$,
\begin{align}
    \bigg\langle (s_{t,j} - h_{t,j}) (s_{t',j} - h_{t',j}) \bigg\rangle_{j} \bigg|_{\hat{C} = 0} 
    &= \Sigma_{t,t'} - \Sigma_{t,t'} - \Sigma_{t',t} + \Sigma_{t,t'} = 0,
\end{align}
while
\begin{align}
    \bigg\langle [(s_{t,j} - h_{t,j})^2 - 1] \bigg\rangle_{j} = 1 + \Sigma_{t,t} - 2 \Sigma_{t,t} + \Sigma_{t,t} - 1 = 0.
\end{align}
Therefore, we conclude at last that the $\hat{C}_{t,t'} = 0$ solution is consistent with the saddle-point equations, and that under this condition we have 
\begin{align}
    C_{t,t'} 
    &= \frac{1}{N} \sum_{j=1}^{N} \langle s_{t,j} s_{t',j} \rangle_{j} 
    \\
    &= \delta_{t,t'} + \Sigma_{t,t'} 
    \\ 
    &= \delta_{t,t'} + \beta^{2} \sum_{k,k'=1}^{K} \Gamma_{k,k'} C_{t-k,t'-k'}
\end{align}
for all interior times $t,t' = 1, \ldots, T$, while the boundary terms for $t = 1, \ldots, T$ and $t' = 1-K, \ldots, 0$ vanish:
\begin{align}
    C_{t,t'} = 0.
\end{align}

More directly, with $\hat{C}_{t,t'} = 0$, the single-site generating function simplifies to 
\begin{align}
    z_{j} &= \mathbb{E}_{\{s_{t,j} \sim \mathcal{N}(0,1)\}} \mathbb{E}_{\mathbf{h}_{j}} \exp\left[\sum_{t=1}^{T} b_{t,j} s_{t,j} + \sum_{t=1}^{T} s_{t,j} h_{t,j} - \frac{1}{2} \sum_{t=1}^{T} h_{t,j}^{2} \right]
    \\ 
    &=  \mathbb{E}_{\{s_{t,j} \sim \mathcal{N}(0,1)\}} \det(\mathbf{I}_{T} + \mathbf{\Sigma})^{-1/2} \exp\left[ \sum_{t=1}^{T} b_{t,j} s_{t,j} + \frac{1}{2} \sum_{t,t'=1}^{T} s_{t,j} (\mathbf{I}_{T} + \mathbf{\Sigma}^{-1})^{-1} s_{t,j} \right]
    \\ 
    &= \exp\left[\frac{1}{2} \sum_{t,t'=1}^{T} b_{t,j} [\mathbf{I}_{T} - (\mathbf{I}_{T} + \mathbf{\Sigma}^{-1})^{-1} ]^{-1} b_{t,j} \right]
    \\ 
    &= \exp\left[\frac{1}{2} \sum_{t,t'=1}^{T} b_{t,j} (\mathbf{I}_{T} + \mathbf{\Sigma}) b_{t,j} \right]. 
\end{align}
Then, we can immediately read off that the site distribution is zero-mean Gaussian, with the covariance claimed above.

\subsection{Summarizing the DMFT}

To summarize, the DMFT for the Gaussian model is given by
\begin{align}
    C_{t,t'} = \delta_{t,t'} + \beta^{2} \sum_{k,k'=1}^{K} \Gamma_{k,k'} C_{t-k,t'-k'}
\end{align}
for $t,t' = 1, \ldots, T$, 
\begin{align}
    C_{t,t'} = 0
\end{align}
for $t = 1, \ldots, T$ and $t' = 1-K, \ldots, 0$, and 
\begin{align}
    C_{t,t'} = \frac{1}{N} \sum_{j=1}^{N} s_{t,j} s_{t',j}
\end{align}
for $t,t' = 1-K, \ldots, 0$. 

Unlike in the Ising case, the equation for $C_{t,t'}$ is now purely linear. We observe that some features of this model cannot carry over to the spherical case, as here $C_{t,t}$ may exceed unity.

\section{Spherical DMFT}\label{app:spherical}

We now consider the spherical model. Using the notation introduced above, we can read off that the disorder average leads to the quenched generating function 
\begin{align}
    Z = \int \prod_{t=1}^{T} d\sigma_{N}(\mathbf{s}_{t}) \, \exp\left[ \sum_{t=1}^{T} \mathbf{b}_{t} \cdot \mathbf{s}_{t} \right] \mathbb{E}_{\mathbf{h}} \prod_{t=1}^{T} \frac{\exp[\mathbf{s}_{t} \cdot \mathbf{h}_{t}]}{\int d\sigma_{N}(\mathbf{u}_{t})\, \exp[\mathbf{u}_{t} \cdot \mathbf{h}_{t}]} 
\end{align}
where the local fields
\begin{align}
    \mathbf{h}_{t} = - \beta \sum_{k=1}^{K} \mathbf{J}_{k} \mathbf{s}_{t-k}
\end{align}
are jointly Gaussian under the distribution of the disorder, and have mean zero at all times. With the specific assumption that
\begin{align}
    \mathbb{E}_{\mathbf{J}}[ (J_{k})_{ij} (J_{k'})_{i'j'}] = \frac{1}{N} \delta_{ii'} \delta_{jj'} \Gamma_{k,k'} ,
\end{align}
their covariance is
\begin{align}
    \mathbb{E}_{\mathbf{J}}[h_{t,j} h_{t',j'}] = \delta_{jj'} \beta^{2} \sum_{k,k'=1}^{K} \Gamma_{k,k'} C_{t-k,t'-k'},
\end{align}
where we define the temporal correlation function of the state: 
\begin{align}
    C_{t,t'} = \frac{1}{N} \mathbf{s}_{t}^{\top}\mathbf{s}_{t'} .
\end{align}
The spherical constraint means that we have 
\begin{align}
    C_{t,t} = 1
\end{align}
at all times. 

\subsection{Introducing order parameters}

As for the other two models, we enforce the definition of $C_{t,t'}$ via Fourier representations of the $\delta$-distribution with Lagrange multipliers $\hat{C}_{t,t'}$, writing
\begin{align}
    1 = \int d\mathbf{C} \int d\hat{\mathbf{C}}\, \exp\left[ - \frac{N}{2} \sum_{t,t'=1-K}^{T} C_{t,t'} \hat{C}_{t,t'} + \frac{1}{2} \sum_{t,t'=1-K}^{T} \hat{C}_{t,t'} \mathbf{s}_{t} \cdot \mathbf{s}_{t'} \right] ,
\end{align}
where in this case we include the diagonal elements of $C_{t,t'}$ and $\hat{C}_{t,t'}$. This yields
\begin{align}
    Z &= \int d\mathbf{C} \int d\hat{\mathbf{C}}\, \exp\left[ - \frac{N}{2} \sum_{t,t'=1-K}^{T} C_{t,t'} \hat{C}_{t,t'} \right]
    \nonumber\\&\quad \times \mathbb{E}_{\mathbf{h}} \int \prod_{t=1}^{T} d\sigma_{N}(\mathbf{s}_{t}) \, \exp\left[ \sum_{t=1}^{T} \mathbf{b}_{t} \cdot \mathbf{s}_{t} + \frac{1}{2} \sum_{t,t'=1-K}^{T} \hat{C}_{t,t'} \mathbf{s}_{t} \cdot \mathbf{s}_{t'} \right] \nonumber\\&\quad \times  \prod_{t=1}^{T} \frac{\exp[\mathbf{s}_{t} \cdot \mathbf{h}_{t}]}{\int d\sigma_{N}(\mathbf{u}_{t})\, \exp[\mathbf{u}_{t} \cdot \mathbf{h}_{t}]} 
\end{align}
Our goal is to factorize 
\begin{align}
    \int \prod_{t=1}^{T} d\sigma_{N}(\mathbf{s}_{t}) \, \exp\left[ \sum_{t=1}^{T} \mathbf{b}_{t} \cdot \mathbf{s}_{t} + \frac{1}{2} \sum_{t,t'=1-K}^{T} \hat{C}_{t,t'} \mathbf{s}_{t} \cdot \mathbf{s}_{t'} \right] \prod_{t=1}^{T} \frac{\exp[\mathbf{s}_{t} \cdot \mathbf{h}_{t}]}{\int d\sigma_{N}(\mathbf{u}_{t})\, \exp[\mathbf{u}_{t} \cdot \mathbf{h}_{t}]} 
\end{align}
over sites. Given that constant multiplicative factors will cancel thanks to the normalization terms, we can re-write this as 
\begin{align}
    &\int \prod_{t=1}^{T} d\mathbf{s}_{t}\, \delta(N - \Vert \mathbf{s}_{t} \Vert^2) \, \exp\left[ \sum_{t=1}^{T} \mathbf{b}_{t} \cdot \mathbf{s}_{t} + \frac{1}{2} \sum_{t,t'=1-K}^{T} \hat{C}_{t,t'} \mathbf{s}_{t} \cdot \mathbf{s}_{t'} \right] \nonumber\\&\qquad \times \prod_{t=1}^{T} \frac{\exp[\mathbf{s}_{t} \cdot \mathbf{h}_{t}]}{\int d\mathbf{u}_{t}\, \delta(N - \Vert \mathbf{u}_{t} \Vert^2) \exp[\mathbf{u}_{t} \cdot \mathbf{h}_{t}]}
    \\ 
    &= \int \prod_{t=1}^{T} d\mathbf{s}_{t}\, \delta(N - \Vert \mathbf{s}_{t} \Vert^2) \, \exp\left[ \sum_{t=1}^{T} \mathbf{b}_{t} \cdot \mathbf{s}_{t} + \frac{1}{2} \sum_{t,t'=1-K}^{T} \hat{C}_{t,t'} \mathbf{s}_{t} \cdot \mathbf{s}_{t'} \right] \nonumber\\&\qquad \times \prod_{t=1}^{T} \frac{\exp[\mathbf{s}_{t} \cdot \mathbf{h}_{t} - \rho \Vert \mathbf{s}_{t} \Vert^2/2]}{\int d\mathbf{u}_{t}\, \delta(N - \Vert \mathbf{u}_{t} \Vert^2) \exp[\mathbf{u}_{t} \cdot \mathbf{h}_{t} - \rho \Vert \mathbf{u}_{t} \Vert^2/2]}
\end{align}
for any positive $\rho$, as the spherical constraint enforces that $\Vert \mathbf{s}_{t} \Vert^2 = \Vert \mathbf{u}_{t} \Vert^2 = N$ for all times $t$. 

At this point, we would like to replace the $\delta$-distributions with their Fourier transforms and interchange the order of integration over the states and Lagrange multipliers, as that would allow us to factor the integral over sites into a product of single-site generating functions. However, the normalizing factor in the denominator means that we cannot proceed so simply. This can be circumvented using the replica trick: we write 
\begin{align}
    \frac{1}{x} = \lim_{n \to 0} x^{n-1}.
\end{align}
More concretely, we introduce $n_{t}-1$ copies of the denominator for the $t$-th timestep. As usual, we will proceed for $n_{t}-1$ a non-negative integer, and then analytically continue the result. Indexing replicas by $a_{t} = 1, \ldots, n_{t}$, the state-dependent term above becomes
\begin{align}
    &\lim_{\{n_{t} \to 0\}} \int \prod_{t=1}^{T} d\mathbf{s}_{t}\, \int \prod_{t=1}^{T} \prod_{a_{t}=1}^{n_{t}-1} d\mathbf{u}_{t}^{a_{t}}\, \left[ \prod_{t=1}^{T} \delta(N - \Vert \mathbf{s}_{t} \Vert^2)  \prod_{a_{t}=1}^{n_{t}} \delta(N - \Vert \mathbf{u}_{t}^{a_{t}}\Vert^{2}) \right] \nonumber\\&\qquad \times \exp\left[ \sum_{t=1}^{T} \mathbf{b}_{t} \cdot \mathbf{s}_{t} + \frac{1}{2} \sum_{t,t'=1-K}^{T} \hat{C}_{t,t'} \mathbf{s}_{t} \cdot \mathbf{s}_{t'} \right] \nonumber\\&\qquad \times \exp\left[\sum_{t=1}^{T} \mathbf{s}_{t} \cdot \mathbf{h}_{t} - \frac{1}{2} \rho \sum_{t=1}^{T} \Vert \mathbf{s}_{t} \Vert^2 + \sum_{t=1}^{T} \sum_{a_{t}=1}^{n_{t}-1} \mathbf{h}_{t} \cdot \mathbf{u}_{t}^{a_{t}} - \frac{1}{2} \rho \sum_{t=1}^{T} \sum_{a_{t}=1}^{n_{t}-1} \Vert \mathbf{u}_{t}^{a_{t}} \Vert^{2} \right] .
\end{align}

We now replace the $\delta$-distributions with their Fourier transforms, introducing Lagrange multipliers $\hat{Q}_{t}$ to enforce the constraints $\Vert \mathbf{s}_{t} \Vert^2 = N$ and $\hat{R}_{t}^{a_{t}}$ to enforce the constraints $\Vert \mathbf{u}_{t}^{a_{t}} \Vert^2 = N$. Interchanging the order of integration, this at last allows us to factor the integrals over sites, yielding
\begin{align}
    Z = \int d\mathbf{C}\, \int d\hat{\mathbf{C}}\, \int d\hat{\mathbf{Q}} \, \int d\hat{\mathbf{R}} \, \exp[N S]
\end{align}
for an action
\begin{align}
    S = - \frac{1}{2} \sum_{t,t'=1-K}^{T} C_{t,t'} \hat{C}_{t,t'} + \frac{1}{2} \sum_{t=1}^{T} \hat{Q}_{t} + \frac{1}{2} \sum_{t=1}^{T} \sum_{a_{t}=1}^{n_{t}-1} \hat{R}_{t}^{a_t} + \sum_{j=1}^{N} \log z_{j} ,
\end{align}
where
\begin{align}
    z_{j} 
    &= \mathbb{E}_{\mathbf{h}_{j}} \int \prod_{t=1}^{T} ds_{t,j}\, \int \prod_{t=1}^{T} \prod_{a_{t}=1}^{n_{t}-1} du_{t,j}^{a_{t}}\, \exp\left[ \sum_{t=1}^{T} b_{t,j} s_{t,j} + \frac{1}{2} \sum_{t,t'=1-K}^{T} \hat{C}_{t,t'} s_{t,j} s_{t',j} + \sum_{t=1}^{T} s_{t,j} h_{t,j}  \right] \nonumber\\&\qquad \times \exp\left[- \frac{1}{2} \sum_{t=1}^{T} (\rho + \hat{Q}_{t}) (s_{t,j})^{2} + \sum_{t=1}^{T} \sum_{a_{t}=1}^{n_{t}-1} h_{t,j} u_{t,j}^{a_{t}} - \frac{1}{2} \sum_{t=1}^{T} \sum_{a_{t}=1}^{n_{t}-1} (\rho + \hat{R}_{t}^{a_{t}}) (u_{t,j}^{a_{t}})^{2} \right] 
\end{align}
are the single-site generating functions. 

\subsection{The saddle-point equations}

Differentiating with respect to $\hat{C}_{t,t'}$ for any two (perhaps equal) times $t,t' = 1-K, \ldots, T$, we of course have
\begin{align}
    C_{t,t'} = \frac{1}{N} \sum_{j=1}^{N} \langle s_{t,j} s_{t',j} \rangle_{j} . 
\end{align}
Here, the self-consistent average is defined via
\begin{align}
    \langle \cdot \rangle_{j} &= \frac{1}{z_{j}} \mathbb{E}_{\mathbf{h}_{j}} \int \prod_{t=1}^{T} ds_{t,j}\, \int \prod_{t=1}^{T} \prod_{a_{t}=1}^{n_{t}-1} du_{t,j}^{a_{t}}\, (\cdot) \exp\left[ \sum_{t=1}^{T} b_{t,j} s_{t,j} + \frac{1}{2} \sum_{t,t'=1-K}^{T} \hat{C}_{t,t'} s_{t,j} s_{t',j} + \sum_{t=1}^{T} s_{t,j} h_{t,j}  \right] \nonumber\\&\qquad \times \exp\left[- \frac{1}{2} \sum_{t=1}^{T} (\rho + \hat{Q}_{t}) (s_{t,j})^{2} + \sum_{t=1}^{T} \sum_{a_{t}=1}^{n_{t}-1} h_{t,j} u_{t,j}^{a_{t}} - \frac{1}{2} \sum_{t=1}^{T} \sum_{a_{t}=1}^{n_{t}-1} (\rho + \hat{R}_{t}^{a_{t}}) (u_{t,j}^{a_{t}})^{2} \right] \bigg|_{b=0} .
\end{align}
As in the Ising and Gaussian cases, one must be careful when considering boundary terms for which $t$ or $t'$ is less than or equal to zero, as the boundary states are fixed by the initial condition. For $t,t' = 1-K, \ldots, 0$, $C_{t,t'}$ is entirely fixed by the boundary condition as
\begin{align}
    C_{t,t'} = \frac{1}{N} \sum_{j=1}^{N} s_{t,j} s_{t',j}.
\end{align}
Now suppose that $t = 1, \ldots, T$ and $t' = 1-K, \ldots, 0$. Then, $s_{t',j}$ is fixed by the initial condition, and we have
\begin{align}
    C_{t,t'} = \frac{1}{N} \sum_{j=1}^{N} \langle s_{t,j} \rangle_{j} s_{t',j}.
\end{align}

Differentiating with respect to $\hat{Q}_{t}$, we have the constraint
\begin{align}
    1 = \frac{1}{N} \sum_{j=1}^{N} \langle (s_{t,j})^2 \rangle_{j}
\end{align}
while differentiating with respect to $\hat{R}_{t}^{a_{t}}$, we have
\begin{align}
    1 = \frac{1}{N} \sum_{j=1}^{N} \langle (u_{t,j}^{a_{t}})^{2} \rangle_{j}
\end{align}

Observing that the auxiliary fields $u_{t,j}^{a_{t}}$ couple to $h_{t,j}$ in the same way as $s_{t,j}$, we make a replica-uniform \emph{Ansatz} for the Lagrange multipliers
\begin{align}
    \hat{R}_{t}^{a_{t}} = \hat{Q}_{t} .
\end{align}
Then, the integrals over $u_{t,j}$ factor, leaving
\begin{align}
    z_{j} 
    &= \mathbb{E}_{\mathbf{h}_{j}} \int \prod_{t=1}^{T} ds_{t,j}\, \exp\left[ \sum_{t=1}^{T} b_{t,j} s_{t,j} + \frac{1}{2} \sum_{t,t'=1-K}^{T} \hat{C}_{t,t'} s_{t,j} s_{t',j} + \sum_{t=1}^{T} s_{t,j} h_{t,j}  \right] \nonumber\\&\qquad \times \exp\left[- \frac{1}{2} \sum_{t=1}^{T} (\rho + \hat{Q}_{t}) (s_{t,j})^{2}  \right] 
    \nonumber\\&\qquad \times \prod_{t=1}^{T} \left\{ \int du_{t,j} \, \exp\left[ h_{t,j} u_{t,j} - \frac{1}{2} (\rho + \hat{Q}_{t}) (u_{t,j})^{2} \right] \right\}^{n_{t}-1}
    \\ 
    &= \mathbb{E}_{\mathbf{h}_{j}} \int \prod_{t=1}^{T} ds_{t,j}\, \exp\left[ \sum_{t=1}^{T} b_{t,j} s_{t,j} + \frac{1}{2} \sum_{t,t'=1-K}^{T} \hat{C}_{t,t'} s_{t,j} s_{t',j} + \sum_{t=1}^{T} s_{t,j} h_{t,j}  \right] \nonumber\\&\qquad \times \exp\left[- \frac{1}{2} \sum_{t=1}^{T} (\rho + \hat{Q}_{t}) (s_{t,j})^{2}  \right] 
    \nonumber\\&\qquad \times \prod_{t=1}^{T} \left\{ \left(\frac{2 \pi}{\rho + \hat{Q}_{t}}\right)^{1/2} \exp\left[ \frac{1}{2} \frac{(h_{t,j})^2}{\rho + \hat{Q}_{t}} \right] \right\}^{n_{t}-1}
    \\ 
    &\overset{n_{t} \to 0}{=} \mathbb{E}_{\mathbf{h}_{j}} \int \prod_{t=1}^{T} ds_{t,j}\, \exp\left[ \sum_{t=1}^{T} b_{t,j} s_{t,j} + \frac{1}{2} \sum_{t,t'=1-K}^{T} \hat{C}_{t,t'} s_{t,j} s_{t',j} + \sum_{t=1}^{T} s_{t,j} h_{t,j}  \right] \nonumber\\&\qquad \times \exp\left[- \frac{1}{2} \sum_{t=1}^{T} (\rho + \hat{Q}_{t}) (s_{t,j})^{2}  \right] 
    \nonumber\\&\qquad \times \prod_{t=1}^{T} \left(\frac{2 \pi}{\rho + \hat{Q}_{t}}\right)^{-1/2} \exp\left[ - \frac{1}{2} \frac{(h_{t,j})^2}{\rho + \hat{Q}_{t}} \right] 
\end{align}
We can then see that $\hat{C}_{t,t'} = 0$ is consistent with normalization. 

To show more carefully that these solutions are consistent, we again turn to the saddle point equation for $\hat{C}_{t,t'}$. As in the Ising and Gaussian cases, this gives 
\begin{align}
    \frac{1}{2} \hat{C}_{t,t} = \frac{1}{N} \sum_{j=1}^{N} \frac{1}{z_{j}} \frac{\partial z_{j}}{\partial C_{t,t}} \bigg|_{b=0}
\end{align}
for equal times and
\begin{align}
    \hat{C}_{t,t'} = \frac{1}{N} \sum_{j=1}^{N} \frac{1}{z_{j}} \frac{\partial z_{j}}{\partial C_{t,t'}} \bigg|_{b=0}
\end{align}
for distinct times $t \neq t'$. Once again, we write 
\begin{align}
    \Sigma_{t,t'} \equiv \mathbb{E}_{\mathbf{J}}[h_{t,j} h_{t',j}] = \beta^{2} \sum_{k,k'=1}^{K} \Gamma_{k,k'} C_{t-k,t'-k'} , 
\end{align}
which gives
\begin{align}
    \frac{1}{z_{j}}\frac{\partial z_{j}}{\partial C_{t,t'}} \bigg|_{b=0} = \beta^{2} \sum_{k,k'=1}^{K} \Gamma_{k,k'} \mathbf{1}\{t+k, t'+k' \geq 1\}  \frac{1}{z_{j}}\frac{\partial z_{j}}{\partial \Sigma_{t+k,t'+k'}} \bigg|_{b=0}. 
\end{align}
Isolating the relevant portion of the single-site generating function, we have for any two distinct times $t,t'$
\begin{align}
    &\frac{\partial}{\partial \Sigma_{t,t'}} \mathbb{E}_{\mathbf{h}_{j}} \exp\left[ \sum_{t=1}^{T} h_{t,j} s_{t,j} + \sum_{t=1}^{T} \sum_{a_{t}=1}^{n_{t}-1} h_{t,j} u_{t,j}^{a_{t}} \right] 
    \\
    &= \mathbb{E}_{\mathbf{h}_{j}} \frac{\partial}{\partial h_{t,j}} \frac{\partial}{\partial h_{t',j}} \exp\left[ \sum_{t=1}^{T} h_{t,j} s_{t,j} + \sum_{t=1}^{T} \sum_{a_{t}=1}^{n_{t}-1} h_{t,j} u_{t,j}^{a_{t}} \right] 
    \\ 
    &= \mathbb{E}_{\mathbf{h}_{j}} \left( s_{t,j} + \sum_{a_{t}=1}^{n_{t}-1} u_{t,j}^{a_{t}} \right) \left( s_{t',j} + \sum_{a_{t'}=1}^{n_{t'}-1} u_{t',j}^{a_{t'}} \right) \exp\left[ \sum_{t=1}^{T} h_{t,j} s_{t,j} + \sum_{t=1}^{T} \sum_{a_{t}=1}^{n_{t}-1} h_{t,j} u_{t,j}^{a_{t}} \right] 
\end{align}
while the equal-time derivatives are
\begin{align}
    &\frac{\partial}{\partial \Sigma_{t,t'}} \mathbb{E}_{\mathbf{h}_{j}} \exp\left[ \sum_{t=1}^{T} h_{t,j} s_{t,j} + \sum_{t=1}^{T} \sum_{a_{t}=1}^{n_{t}-1} h_{t,j} u_{t,j}^{a_{t}} \right] 
    \\
    &= \frac{1}{2} \mathbb{E}_{\mathbf{h}_{j}} \frac{\partial^2}{\partial h_{t,j}^2} \exp\left[ \sum_{t=1}^{T} h_{t,j} s_{t,j} + \sum_{t=1}^{T} \sum_{a_{t}=1}^{n_{t}-1} h_{t,j} u_{t,j}^{a_{t}} \right] 
    \\ 
    &= \frac{1}{2} \mathbb{E}_{\mathbf{h}_{j}} \left( s_{t,j} + \sum_{a_{t}=1}^{n_{t}-1} u_{t,j}^{a_{t}} \right)^2 \exp\left[ \sum_{t=1}^{T} h_{t,j} s_{t,j} + \sum_{t=1}^{T} \sum_{a_{t}=1}^{n_{t}-1} h_{t,j} u_{t,j}^{a_{t}} \right] .
\end{align}
Using the definition of the single-site average, we then have
\begin{align}
    \frac{1}{z_{j}} \frac{\partial z_{j}}{\partial \Sigma_{t,t'}} \bigg|_{b=0} 
    &= \left\langle \left( s_{t,j} + \sum_{a_{t}=1}^{n_{t}-1} u_{t,j}^{a_{t}} \right) \left( s_{t',j} + \sum_{a_{t'}=1}^{n_{t'}-1} u_{t',j}^{a_{t'}} \right) \right\rangle_{j} 
\end{align}
for any two distinct times $t,t'$, while
\begin{align}
    \frac{1}{z_{j}} \frac{\partial z_{j}}{\partial \Sigma_{t,t}} \bigg|_{b=0} 
    &= \frac{1}{2} \left\langle \left( s_{t,j} + \sum_{a_{t}=1}^{n_{t}-1} u_{t,j}^{a_{t}} \right)^2 \right\rangle_{j} 
\end{align}
Importantly, in these expectations $t$ and $t'$ are constrained to be strictly greater than zero. 

We can evaluate these expectations using the fact that the single-site distribution of the fields at positive times is in fact Gaussian. We condition on the fields $h_{t,j}$, set $b_{t,j} = 0$, and consider the joint density of the fields $s_{t,j}$ and $u_{t,j}^{a_{t}}$. Defining the matrix
\begin{align}
    A_{t,t'} = (\rho + \hat{Q}_{t}) \delta_{t,t'} - \hat{C}_{t,t'} ,
\end{align}
for $t,t' = 1, \ldots, T$ and the vector
\begin{align}
    c_{t,j} = \sum_{t'=1-K}^{0} \hat{C}_{t,t'} s_{t',j}
\end{align}
and using the fact that terms depending on $\hat{C}_{t,t'}$ for both $t,t' \leq 0$ will drop as they contribute a constant, we have that their joint density is given up to normalization as
\begin{align}
    & \exp\left[ -\frac{1}{2} \sum_{t,t'=1}^{T} A_{t,t'} s_{t,j} s_{t',j} + \sum_{t=1}^{T} s_{t,j} (h_{t,j} + c_{t,j}) + \sum_{t=1}^{T} \sum_{a_{t}=1}^{n_{t}-1} h_{t,j} u_{t,j}^{a_{t}} - \frac{1}{2} \sum_{t=1}^{T} \sum_{a_{t}=1}^{n_{t}-1} (\rho + \hat{R}_{t}^{a_{t}}) (u_{t,j}^{a_{t}})^{2} \right]
    \\ 
    &= \exp\left[ -\frac{1}{2} \sum_{t,t'=1}^{T} A_{t,t'} s_{t,j} s_{t',j} + \sum_{t=1}^{T} s_{t,j} (h_{t,j} + c_{t,j}) \right] \prod_{t=1}^{T} \prod_{a_{t}=1}^{n_{t}-1} \exp\left[ - \frac{1}{2} (\rho + \hat{R}_{t}^{a_{t}}) (u_{t,j}^{a_{t}})^{2} - h_{t,j} u_{t,j}^{a_{t}} \right] .
\end{align}
We can then see that the replica fields $u_{t,j}^{a_{t}}$ are independent from the fields $s_{t,j}$ and are independent across replicas and time, with variance
\begin{align}
    \var[ u_{t,j}^{a_{t}} | \mathbf{h}_{j} ] = \frac{1}{\rho + \hat{R}_{t}^{a_{t}}}
\end{align}
and mean
\begin{align}
    \mathbb{E}[ u_{t,j}^{a_{t}} | \mathbf{h}_{j} ] = \frac{1}{\rho + \hat{R}_{t}^{a_{t}}} h_{t,j}.
\end{align}
Similarly, the fields $s_{t,j}$ are Gaussian, with covariance
\begin{align}
    \cov[s_{t,j}, s_{t',j} | \mathbf{h}_{j} ] = (A^{-1})_{t,t'}
\end{align}
and mean
\begin{align}
    \mathbb{E}[s_{t,j} | \mathbf{h}_{j} ] = \sum_{t'=1}^{T} (A^{-1})_{t,t'} (h_{t,j} + c_{t,j}).
\end{align}

Though we can compute the required correlators using these full expressions, we recall that our goal is to show that they vanish provided that $\hat{C}_{t,t'} = 0$ and $\hat{R}_{t}^{a_{t}} = \hat{Q}_{t}$. We will therefore assume that $\hat{C}_{t,t'} = 0$, which yields the drastic simplification
\begin{align}
    \cov[s_{t,j}, s_{t',j} | \mathbf{h}_{j} ] \bigg|_{\hat{C}_{t,t'} = 0} = \frac{1}{\rho + \hat{Q}_{t}} \delta_{t,t'},
\end{align}
\begin{align}
    \mathbb{E}[s_{t,j} | \mathbf{h}_{j} ]  \bigg|_{\hat{C}_{t,t'} = 0} = \frac{1}{\rho + \hat{Q}_{t}} h_{t,j} .
\end{align}
Using these facts, we can easily compute the unequal-time correlators as
\begin{align}
    \frac{1}{z_{j}} \frac{\partial z_{j}}{\partial \Sigma_{t,t'}} \bigg|_{b=0}  \bigg|_{\hat{C}_{t,t'} = 0}
    &= \left\langle \left( s_{t,j} + \sum_{a_{t}=1}^{n_{t}-1} u_{t,j}^{a_{t}} \right) \left( s_{t',j} + \sum_{a_{t'}=1}^{n_{t'}-1} u_{t',j}^{a_{t'}} \right) \right\rangle_{j}  \bigg|_{\hat{C}_{t,t'} = 0}
    \\ 
    &= \langle s_{t,j} s_{t',j} \rangle_{j} + \sum_{a_{t}=1}^{n_{t}-1} \langle u_{t,j}^{a_{t}} s_{t',j} \rangle_{j} + \sum_{a_{t'}=1}^{n_{t'}-1} \langle s_{t,j} u_{t',j}^{a_{t'}} \rangle_{j} + \sum_{a_{t}=1}^{n_{t}-1} \sum_{a_{t'}=1}^{n_{t'}-1} \langle u_{t,j}^{a_{t}} u_{t',j}^{a_{t'}} \rangle_{j}  \bigg|_{\hat{C}_{t,t'} = 0}
    \\ 
    &= \frac{1}{(\rho + \hat{Q}_{t}) (\rho + \hat{Q}_{t'})} \mathbb{E}_{\mathbf{h}_{j}}[h_{t,j} h_{t',j}] 
    \nonumber\\&\quad + \sum_{a_{t}=1}^{n_{t}-1} \frac{1}{(\rho + \hat{Q}_{t'}) (\rho + \hat{R}_{t}^{a_{t}})} \mathbb{E}_{\mathbf{h}_{j}}[h_{t,j} h_{t',j}]
    \nonumber\\&\quad + \sum_{a_{t'}=1}^{n_{t'}-1} \frac{1}{(\rho + \hat{Q}_{t}) (\rho + \hat{R}_{t'}^{a_{t'}})} \mathbb{E}_{\mathbf{h}_{j}}[h_{t,j} h_{t',j}]
    \nonumber\\&\quad + \sum_{a_{t}=1}^{n_{t}-1} \sum_{a_{t'}=1}^{n_{t'}-1}  \frac{1}{(\rho + \hat{R}_{t}^{a_{t}}) (\rho + \hat{R}_{t'}^{a_{t'}})} \mathbb{E}_{\mathbf{h}_{j}}[h_{t,j} h_{t',j}] 
    \\ 
    &= \left( \frac{1}{\rho + \hat{Q}_{t}} + \sum_{a_{t}=1}^{n_{t}-1} \frac{1}{\rho + \hat{R}_{t}^{a_{t}}} \right) \left( \frac{1}{\rho + \hat{Q}_{t'}} + \sum_{a_{t'}=1}^{n_{t'}-1} \frac{1}{\rho + \hat{R}_{t'}^{a_{t'}}} \right) \Sigma_{t,t'}
\end{align}
for any two distinct times $t,t'$, while the equal-time correlators are
\begin{align}
    2 \frac{1}{z_{j}} \frac{\partial z_{j}}{\partial \Sigma_{t,t}} \bigg|_{b=0} 
    &= \left\langle \left( s_{t,j} + \sum_{a_{t}=1}^{n_{t}-1} u_{t,j}^{a_{t}} \right)^2 \right\rangle_{j} 
    \\ 
    &= \langle s_{t,j} s_{t,j} \rangle_{j} + 2 \sum_{a_{t}=1}^{n_{t}-1} \langle s_{t,j} u_{t,j}^{a_{t}} \rangle_{j} + \sum_{a_{t}=1}^{n_{t}-1} \sum_{a_{t}=1}^{n_{t}-1} \langle u_{t,j}^{a_{t}} u_{t,j}^{a_{t}} \rangle_{j}  \bigg|_{\hat{C}_{t,t'} = 0}
    \\ 
    &= \frac{1}{\rho + \hat{Q}_{t}} + \frac{1}{(\rho + \hat{Q}_{t})^2} \mathbb{E}_{\mathbf{h}_{j}}[h_{t,j}^2] 
    \nonumber\\&\quad + 2 \sum_{a_{t}=1}^{n_{t}-1} \frac{1}{(\rho + \hat{Q}_{t}) (\rho + \hat{R}_{t}^{a_{t}})} \mathbb{E}_{\mathbf{h}_{j}}[h_{t,j}^2]
    \nonumber\\&\quad + \sum_{a_{t}=1}^{n_{t}-1} \sum_{a_{t}'=1}^{n_{t}-1} \left(\frac{1}{\rho + \hat{R}_{t}^{a_{t}}} \delta_{a_{t} a_{t}'} + \frac{1}{(\rho + \hat{R}_{t}^{a_{t}}) (\rho + \hat{R}_{t}^{a_{t}'})} \mathbb{E}_{\mathbf{h}_{j}}[h_{t,j}^2] \right)
    \\ 
    &= \frac{1}{\rho + \hat{Q}_{t}} + \sum_{a_{t}=1}^{n_{t}-1} \frac{1}{1 + \hat{R}_{t}^{a_{t}}} 
    \nonumber\\&\quad + \left( \frac{1}{\rho + \hat{Q}_{t}} + \sum_{a_{t}=1}^{n_{t}-1} \frac{1}{\rho + \hat{R}_{t}^{a_{t}}} \right) \left( \frac{1}{\rho + \hat{Q}_{t'}} + \sum_{a_{t'}=1}^{n_{t'}-1} \frac{1}{\rho + \hat{R}_{t'}^{a_{t'}}} \right) \Sigma_{t,t}
\end{align}
From this, we can see immediately that the replica-uniform \emph{Ansatz} 
\begin{align}
    \hat{R}_{t}^{a_{t}} = \hat{Q}_{t}
\end{align}
gives 
\begin{align}
    \frac{1}{z_{j}} \frac{\partial z_{j}}{\partial \Sigma_{t,t'}} \bigg|_{b=0}  \bigg|_{\hat{C}_{t,t'} = 0} = n_{t} n_{t'} \frac{1}{(\rho + \hat{Q}_{t}) (\rho + \hat{Q}_{t'})} \Sigma_{t,t'} 
\end{align}
and
\begin{align}
    \frac{1}{z_{j}} \frac{\partial z_{j}}{\partial \Sigma_{t,t}} \bigg|_{b=0}  \bigg|_{\hat{C}_{t,t'} = 0} = n_{t} \frac{1}{\rho + \hat{Q}_{t}} + n_{t}^2 \frac{1}{(\rho + \hat{Q}_{t})^2} \Sigma_{t,t}, 
\end{align}
which vanish as $n_{t} \downarrow 0$, meaning that the $\hat{C}_{t,t'} = 0$ solution is self-consistent in this case.

With $\hat{C}_{t,t'} = 0$, the single-site generating function becomes
\begin{align}
    z_{j} 
    &= \mathbb{E}_{\mathbf{h}_{j}} \int \prod_{t=1}^{T} ds_{t,j}\, \exp\left[ \sum_{t=1}^{T} b_{t,j} s_{t,j} - \frac{1}{2} \sum_{t=1}^{T} (\rho + \hat{Q}_{t}) (s_{t,j})^{2}  + \sum_{t=1}^{T} s_{t,j} h_{t,j}  \right] 
    \nonumber\\&\qquad \times \prod_{t=1}^{T} \left(\frac{2 \pi}{\rho + \hat{Q}_{t}}\right)^{-1/2} \exp\left[ - \frac{1}{2} \frac{(h_{t,j})^2}{\rho + \hat{Q}_{t}} \right] . 
\end{align}
Defining the diagonal matrix $\mathbf{Q}$ by
\begin{align}
    Q_{t,t} = \rho + \hat{Q}_{t},
\end{align}
we have
\begin{align}
    z_{j} 
    &= (\det \mathbf{Q})^{+1/2} \int \prod_{t=1}^{T} \frac{ds_{t,j}}{(2\pi)^{1/2}} \, \exp\left[ \sum_{t=1}^{T} b_{t,j} s_{t,j} - \frac{1}{2} \sum_{t=1}^{T} Q_{t,t} (s_{t,j})^{2} \right] \nonumber\\&\quad\qquad\times (\det \mathbf{\Sigma})^{-1/2} \int \prod_{t=1}^{T} \frac{dh_{t,j}}{(2\pi)^{1/2}} \exp\left[ - \frac{1}{2} \sum_{t,t'=1}^{T} (\Sigma^{-1} + Q^{-1})_{t,t'} h_{t,j} h_{t',j} + \sum_{t=1}^{T} h_{t,j} s_{t,j} \right]
    \\ 
    &= \det(\mathbf{Q}^{-1} + \mathbf{Q}^{-1}\mathbf{\Sigma}\mathbf{Q}^{-1})^{-1/2} 
    \nonumber\\&\quad \times \int \prod_{t=1}^{T} \frac{ds_{t,j}}{(2\pi)^{1/2}} \, \exp\left[ \sum_{t=1}^{T} b_{t,j} s_{t,j} - \frac{1}{2} \sum_{t,t'=1}^{T} [\mathbf{Q} - (\mathbf{\Sigma}^{-1} + \mathbf{Q}^{-1})^{-1}]_{t,t'} s_{t,j} s_{t',j} \right]
    \\ 
    &= \exp\left[ \frac{1}{2} \sum_{t,t'=1}^{T} [\mathbf{Q}^{-1} (\mathbf{Q} + \mathbf{\Sigma}) \mathbf{Q}^{-1}]_{t,t'} b_{t,j} b_{t',j} \right], 
\end{align}
where we observe that the Woodbury identity implies that 
\begin{align}
    \mathbf{Q} - (\mathbf{\Sigma}^{-1} + \mathbf{Q}^{-1})^{-1}
    &= \mathbf{Q} (\mathbf{Q} + \mathbf{\Sigma})^{-1} \mathbf{Q} .
\end{align}
This gives 
\begin{align}
    C_{t,t'} = [\mathbf{Q}^{-1} (\mathbf{Q} + \mathbf{\Sigma}) \mathbf{Q}^{-1}]_{t,t'} 
\end{align}
for $t,t' = 1, \ldots, T$, which as $Q$ is diagonal reduces to
\begin{align}
    C_{t,t'} = \frac{1}{Q_{t} Q_{t'}} \left( Q_{t} \delta_{t,t'} + \beta^{2} \sum_{k,k'=1}^{K} \Gamma_{k,k'} C_{t-k,t'-k'} \right) .
\end{align}
To determine $Q_{t}$, we use the self-consistency condition
\begin{align}
    C_{t,t} = 1 ;
\end{align}
as $\rho$ is arbitrary we can absorb it into the re-definition. Moreover, we have $\langle s_{t,j} \rangle_{j} = 0$, hence $C_{t,t'} = 0$ if $t>0$ and $t' \leq 0$.

\subsection{Summarizing the DMFT}

To summarize, for the spherical model we have obtained a DMFT in terms of the two-point functions
\begin{align}
    C_{t,t'} = \frac{1}{N} \mathbf{s}_{t} \cdot \mathbf{s}_{t'}
\end{align}
and a set of positive scalars $Q_{t}$, where $C_{t,t'}$ satisfies the recursive equation
\begin{align}
    C_{t,t'} = \frac{1}{Q_{t} Q_{t'}} \left( Q_{t} \delta_{t,t'} + \beta^{2} \sum_{k,k'=1}^{K} \Gamma_{k,k'} C_{t-k,t'-k'} \right) 
\end{align}
and $Q_{t}$ is determined by the self-consistency condition 
\begin{align}
    C_{t,t} = 1.
\end{align}
These DMFT equations are naturally solved via fixed point iteration. For a fixed $Q_{t}$, this is a linear recurrence for $C_{t,t'}$ which can be solved forward in time. Then, we can compute $Q_{t}$ from the given $C_{t,t'}$ using the constraint $C_{t,t} = 1$. This alternating procedure can be iterated until convergence. 

\section{Numerical methods}

All simulations were done using MATLAB R2024b (The MathWorks, Natick, MA, USA), and were run on a desktop workstation equipped with an Intel Xeon(R) w5-3525 processor and 128 GB of RAM. They were not compute-intensive, requiring well under one CPU-hour. For the spherical model, we used Sungkyu Jung's freely available implementation of \citet{wood1994vmf}'s algorithm for sampling from the von Mises-Fisher distribution, available at \url{https://www.stat.pitt.edu/sungkyu/oldSoftwarePage.html}.

\end{document}